\documentclass[11pt]{article}

\usepackage[left=1.3in,right=1.3in,top=1in,bottom=1in]{geometry}

\usepackage{amsmath,amssymb,amsfonts,amsthm}
\allowdisplaybreaks

\usepackage{graphicx}
\usepackage{xcolor}
\usepackage{booktabs}
\usepackage[font=small]{caption}
\usepackage{subcaption}
\usepackage{adjustbox}
\usepackage{makecell}
\usepackage{array}
\usepackage{threeparttable}
\usepackage{multirow}

\usepackage{algorithm}
\usepackage{algpseudocode}

\usepackage[colorlinks=true,urlcolor=blue,citecolor=red,linkcolor=blue]{hyperref}

\newtheorem{theorem}{Theorem}[section]

\newtheorem{lemma}[theorem]{Lemma}
\newtheorem{proposition}[theorem]{Proposition}

\theoremstyle{definition}

\newtheorem{remark}[theorem]{Remark}

\title{Transformed $\ell_1$ Gradient Regularization for Image Denoising}

\author{
Nabiha Choudhury\textsuperscript{1}
\and
Jianqing Jia\textsuperscript{2}
\and
Yifei Lou\textsuperscript{1,2,*}
}

\date{}

\begin{document}

\maketitle

\vspace{-1.5em}

\begin{center}
\small
\textsuperscript{1}Department of Mathematics, University of North Carolina at Chapel Hill, Chapel Hill, NC 27599, USA

\textsuperscript{2}School of Data and Information Sciences, University of North Carolina at Chapel Hill, Chapel Hill, NC 27599, USA

\end{center}

\begin{abstract}
Total variation (TV) regularization is a classical edge-preserving technique widely used across image recovery and reconstruction problems; however, its convex $\ell_1$ gradient penalty tends to over-shrink large gradients, producing staircase artifacts and contrast loss. We propose a gradient-based regularization using the Transformed $\ell_1$ (TL1) penalty and apply it to image denoising. The TL1 penalty asymptotically interpolates between $\ell_1$ and the $\ell_0$ pseudo-norm, offering a principled alternative to TV that better preserves sharp edges and piecewise-smooth regions. Moreover, TL1 admits a tractable proximal operator, enabling an efficient algorithm based on a proximal splitting scheme with subproblems solved by the Alternating Direction Method of Multipliers (ADMM). 
The weak convexity of TL1 guarantees global convergence of the proximal iterates to a stationary point under mild conditions. Numerical experiments on image denoising demonstrate that the proposed method effectively preserves sharp edges, local contrast, and piecewise-smooth structures, outperforming other gradient-based approaches.
\end{abstract}

\newcommand\keywordsname{Key words}

\newenvironment{keywords}{
     \vspace{.05in}\footnotesize
     \parindent .0in
     {\upshape\bfseries \keywordsname. }\ignorespaces
     }
     {\par\vspace{.1in}}

\makeatletter
\newcommand\blfootnote[1]{%
  \begingroup
  \renewcommand\thefootnote{}\footnote{#1}%
  \addtocounter{footnote}{-1}%
  \endgroup
}
\makeatother

\blfootnote{Emails: \texttt{chnabi@unc.edu}, \texttt{jqjia@unc.edu}, \texttt{yflou@unc.edu}. }


\section{Introduction}

Image denoising aims to recover an unknown clean image from a noisy observation. 
In particular, the observed image is often written as 
\begin{equation*}
    \mathbf{f} = \mathbf{u} + \boldsymbol{\varepsilon},
\end{equation*}
where $\mathbf{u}$ denotes the clean image, $\mathbf{f}$ is the noisy observation, and $\boldsymbol{\varepsilon}$ is assumed to be additive Gaussian noise.
Since the recovery of $\mathbf{u}$ from $\mathbf{f}$ is generally ill-posed, variational regularization methods \cite{rudin1992nonlinear,lysaker2003noise,lou2010image,YifeiAni-iso2015,mcpYou2018,wang2022minimizing} have been widely used. 
This leads to the model
\begin{equation*}
    \min_{\mathbf{u}} \; \frac{\mu}{2}\|\mathbf{u}-\mathbf{f}\|_2^2 + R(\mathbf{u}),
\end{equation*}
where the first term enforces consistency with the noisy observation, the regularization $R(\mathbf{u})$ encodes prior structural information, and $\mu>0$ controls the balance between the two.

A fundamental prior in image denoising is gradient sparsity. 
Natural images are often approximately piecewise smooth: most regions vary slowly, while edges appear as sharp discontinuities. 
Therefore, sparsity-promoting regularization on image gradients can effectively suppress noise while preserving important structural features. Total Variation (TV) \cite{rudin1992nonlinear} is a classical gradient-based regularization of this kind, penalizing the $\ell_1$ norm of the image gradient to promote sparsity. 
Despite its success, the convex $\ell_1$ formulation is known to produce staircase artifacts and contrast loss \cite{lysaker2003noise}, as the $\ell_1$ penalty imposes equal shrinkage on all gradient components regardless of magnitude, introducing a systematic 
bias that over-penalizes large gradients corresponding to 
image edges. This is further compounded by the fact that the $\ell_1$
norm is only a convex relaxation of the discontinuous $\ell_0$ pseudo-norm, which explicitly counts nonzero entries and promotes sparsity more aggressively. 
Since direct $\ell_0$ optimization is combinatorially intractable, nonconvex regularizers that better approximate $\ell_0$ sparsity while remaining tractable have attracted considerable attention.

In the direction of sparse recovery, a broad class of nonconvex regularizers has been proposed to reduce the bias of the convex $\ell_1$ relaxation, including Smoothly Clipped Absolute Deviation (SCAD) \cite{fan2001variable}, Minimax Concave Penalty (MCP)  \cite{zhang2010nearly}, and capped-$\ell_1$ \cite{zhang2008multi,lou2016point}. A related family of concave penalties imposes weaker shrinkage on large components, such as
 the $\ell_p$ quasi-norm $(0<p<1)$ \cite{frank1993statistical,xu2012l_,chen2016computing} and error function \cite{guo2021novel}. Other nonconvex sparsity-promoting regularizations include $\ell_1$--$\ell_2$ \cite{yin2015minimization,lou2015computing,lou2018fast},  $\ell_1/\ell_2$ \cite{rahimi2019scale,wang2020accelerated}, and $(\ell_1/\ell_2)^2$ \cite{jia2025sparse,jia2026signal}.

These nonconvex penalties can be applied to image gradients to promote sparse gradient structures, offering improvements over TV in edge and contrast preservation. For example, 
Lou et al.~\cite{YifeiAni-iso2015} proposed the $\ell_1$--$\ell_2$ regularization on the gradient, interpretable as a weighted difference between anisotropic and isotropic TV. MCP and a hyperbolic tangent penalty were applied to image restoration in \cite{mcpYou2018} and \cite{li2023nonconvex}, respectively; their concavity reduces shrinkage bias on large gradient components while still encouraging sparsity. 
The scale-invariant $\ell_1/\ell_2$ regularization on the gradient was investigated in \cite{wang2021limited,wang2022minimizing}, aiming to alleviate the contrast loss inherent to TV. 
Despite these advantages, the nonsmooth and nonconvex nature of such models often entails heavy computational costs, and convergence guarantees, when available, typically rely on strong or restrictive assumptions.

Motivated by these challenges, we propose to apply the Transformed 
$\ell_1$ (TL1) penalty~\cite{zhang2017minimization,zhang2018minimization} 
to the image gradient as a regularizer for image denoising. 
Originally developed for sparse recovery in the compressed sensing 
setting~\cite{zhang2017minimization,zhang2018minimization}, TL1 
has since been applied to promote low-rankness in matrix 
recovery problems~\cite{zhang2017transformed,zhao2025noisy,zhao2026transformed}. To the best of our knowledge, 
its application to image gradient regularization has not been explored.  For a scalar variable $t\in \mathbb{R}$, the TL1 penalty is defined by
\begin{equation}
    \phi_a(t) = \frac{(a+1)|t|}{a+|t|}, \qquad a>0.
    \label{eq:scalar_tl1}
\end{equation}
TL1 is attractive for three reasons. First, it asymptotically interpolates between the $\ell_1$ norm and $\ell_0$ pseudo-norm: 
\begin{equation*}
    \lim_{a \to \infty} \phi_a(t) = |t|,
    \qquad
    \lim_{a \to 0^+} \frac{\phi_a(t)}{a+1} = \mathbf{1}_{[t \neq 0]},
    \label{eq:tl1_interpolation}
\end{equation*}
where $\mathbf{1}_{[t \neq 0]}$ denotes the $\ell_0$ pseudo-norm.
As $a$ decreases, $\phi_a$ promotes stronger sparsity approaching 
the $\ell_0$ pseudo-norm, while for large $a$ it approaches the 
$\ell_1$ norm.
This makes TL1  particularly suitable for gradient regularization, where small oscillatory gradients caused by noise should be suppressed while large gradients corresponding to image edges should be preserved with less bias. Second, TL1 admits a closed-form proximal mapping \cite{zhang2017minimization}, enabling an efficient algorithm based on a proximal splitting scheme.  Third, TL1 is weakly convex, which allows us to establish convergence of the proposed algorithm to a stationary point under mild conditions. 

Applying TL1 componentwise to the discrete image gradient leads to the denoising model
\begin{equation}\label{eq:regularized-denoising}
    \min_\mathbf{u} \; \frac{\mu}{2}\|\mathbf{u}-\mathbf{f}\|_2^2 + G(D \mathbf{u}),
\end{equation}
where $D$ denotes the discrete gradient operator (defined 
precisely in Section~\ref{subsec:model}) and $G$ applies 
the TL1 penalty componentwise to $D\mathbf{u}$.
To solve the resulting nonsmooth and nonconvex problem 
\eqref{eq:regularized-denoising}, we introduce an auxiliary 
variable $\mathbf{d} = D\mathbf{u}$ and reformulate it as 
a constrained problem, which is then solved by an alternating 
minimization scheme. A proximal regularization term is added 
to the $\mathbf{d}$-subproblem at each iteration to stabilize 
the nonconvex update. This is necessary because a single-loop 
scheme applied directly to the nonconvex constrained problem 
cannot control the constraint residual, due to the nontrivial 
null space of the discrete gradient operator $D$. At each outer iteration, the resulting subproblem is solved 
by an Alternating Direction Method of Multipliers (ADMM) 
\cite{boyd2011distributed} as the inner solver. Within ADMM, 
the $\mathbf{u}$- and $\mathbf{d}$-subproblems are decoupled: 
under periodic boundary conditions, the image update is 
computed efficiently via the Fast Fourier Transform (FFT), 
while the gradient update is separable and admits a 
closed-form TL1 proximal operator. Numerical experiments on benchmark images 
demonstrate that the proposed TL1 model is competitive with and often outperforms 
TV~\cite{rudin1992nonlinear}, $\ell_1$--$\ell_2$~\cite{YifeiAni-iso2015}, 
MCP~\cite{mcpYou2018}, and a combination of a logarithmic penalty 
with high-order TV (LOG+TV)~\cite{FAN202585} in terms of PSNR and 
SSIM across multiple noise levels, with particularly strong 
performance on images with sparse, high-contrast gradient structures. 
The main contributions of this paper are summarized as follows.
\begin{itemize}
    \item \textbf{Model.} We propose a TL1 gradient-regularized 
    denoising model as a principled nonconvex alternative to TV 
    regularization. The TL1 penalty uniquely combines three 
    favorable properties: asymptotic interpolation between the 
$\ell_1$ norm and the $\ell_0$ pseudo-norm, weak convexity, and
    a closed-form proximal operator. 
    These properties together enable both an efficient algorithm 
    and a verifiable convergence guarantee, making TL1 a 
    particularly suitable regularizer for gradient-based 
    image denoising.

    \item \textbf{Algorithm.} We develop an efficient algorithm based 
on a proximal splitting scheme, in which each outer iteration 
adds a proximal regularization term to the gradient variable 
and the resulting subproblem is solved by ADMM, which requires 
only an FFT-based image update and a componentwise closed-form 
TL1 proximal operator.

  \item \textbf{Convergence.} We establish convergence of the 
whole sequence of iterates to a stationary point of the TL1 
denoising model using the Kurdyka--\L{}ojasiewicz (KL) 
framework~\cite{attouch2013convergence}. The convergence 
condition links the degree of nonconvexity of TL1 explicitly 
to the required proximal regularization strength.
\end{itemize}

The rest of this paper is organized as follows. Section~2 introduces the proposed TL1 gradient-regularized denoising model and derives the algorithm. Section~3 presents the convergence analysis. 
Section~4 reports numerical experiments and comparisons with related regularization methods. Section~5 concludes the paper.

\section{The Proposed Approach}
\label{sec:problem}

This section presents the proposed TL1 gradient-regularized 
denoising model and the algorithm used to solve it. Specifically,
Section~\ref{subsec:model} introduces the discrete gradient 
operator and formulates the TL1 denoising model. A proximal regularization 
term is added to the gradient variable to stabilize the 
nonconvex update, leading to the outer proximal scheme. 
Section~\ref{subsec:tl1_properties} elaborates on two of 
these properties that are central to the algorithm design 
and convergence analysis: weak convexity with an explicit 
modulus, which supports the convergence analysis in 
Section~\ref{sec:convergence}, and a closed-form 
proximal operator, which enables an efficient $\mathbf{d}$-update 
in the ADMM inner solver.
Section~\ref{subsec:admm} derives the ADMM inner solver, 
detailing the FFT-based image update, the componentwise 
closed-form TL1 proximal update for the gradient variable, 
and the dual variable update.

\subsection{TL1 Denoising Model}\label{subsec:model}
Let $U \in \mathbb{R}^{N\times M}$ denote a two-dimensional image of size $N\times M$, and let 
$\mathbf{u}=\operatorname{vec}(U)\in\mathbb{R}^{n}$, where $n=NM$, be its vectorized representation. 
Under periodic boundary conditions, the horizontal and vertical forward differences are defined by
\begin{equation*}
    (D_x U)_{i,j}=U_{i,j+1}-U_{i,j},
    \qquad
    (D_y U)_{i,j}=U_{i+1,j}-U_{i,j}.
\end{equation*}
We use the same notation $D_x$ and $D_y$ for the corresponding matrix operators acting on the vectorized image $\mathbf{u}$. 
The discrete gradient operator is then defined as
\begin{equation*}
    D\mathbf{u}
    =
    \begin{pmatrix}
        D_x\mathbf{u}\\
        D_y\mathbf{u}
    \end{pmatrix}
    \in \mathbb{R}^{2n},
    \label{eq:gradient}
\end{equation*}
with adjoint $D^\top = (D_x^\top, D_y^\top)$.

Let $\phi_a$ be the scalar TL1 penalty recalled in \eqref{eq:scalar_tl1} of the introduction. 
For $\mathbf{d}=(\mathbf{d}_x,\mathbf{d}_y)\in\mathbb{R}^{2n}$ with
$\mathbf{d}_x,\mathbf{d}_y\in\mathbb{R}^n$, define
\begin{align}
    G(\mathbf{d}) := \sum_{\ell=1}^n \left[\phi_a((\mathbf{d}_x)_\ell) + \phi_a((\mathbf{d}_y)_\ell)\right].
    \label{eq:G_def}
\end{align}
Using the TL1 gradient penalty $G$ defined in \eqref{eq:G_def}, the proposed TL1 gradient-regularized denoising model is
\begin{equation}
    \min_{\mathbf{u}\in\mathbb{R}^n}
    \;
    F(\mathbf{u})+G(D\mathbf{u}),
    \qquad
    F(\mathbf{u})
    :=
    \frac{\mu}{2}\|\mathbf{u}-\mathbf{f}\|_2^2,
    \label{eq:unconstrained}
\end{equation}
where $\mathbf{f}$ is the observed noisy image and $\mu > 0$ 
is a parameter that balances data fidelity against regularization.
Equivalently, by introducing auxiliary gradient variables
\begin{equation*}
    \mathbf{d}_x=D_x\mathbf{u},
    \qquad
    \mathbf{d}_y=D_y\mathbf{u},
    \label{eq:aux_vars}
\end{equation*}
we obtain the constrained formulation
\begin{equation}
\begin{aligned}
    \min_{\mathbf{u},\mathbf{d}_x,\mathbf{d}_y}
    \quad&
    F(\mathbf{u})+G(\mathbf{d}_x,\mathbf{d}_y)\\
    \text{s.t.}\quad&
    \mathbf{d}_x=D_x\mathbf{u},
    \qquad
    \mathbf{d}_y=D_y\mathbf{u}.
\end{aligned}
\label{eq:constrained}
\end{equation}

Since the nonsmooth and nonconvex component of 
\eqref{eq:constrained} appears only in the gradient variable 
$\mathbf{d}=(\mathbf{d}_x,\mathbf{d}_y)$, we add a proximal 
regularization term to the $\mathbf{d}$-block and consider 
the following iterative scheme:
\begin{equation}
(\mathbf{u}^{k+1},\mathbf{d}^{k+1})
\in
\arg\min_{\mathbf{u},\mathbf{d}:\,\mathbf{d}=D\mathbf{u}}
\left\{
F(\mathbf{u})+G(\mathbf{d})
+
\frac{\eta}{2}\|\mathbf{d}-\mathbf{d}^k\|_2^2
\right\},
\tag{$\mathcal{P}^k_\eta$}
\label{eq:proximal_scheme}
\end{equation}
where $\eta>0$ is the proximal regularization parameter, 
and the iteration is initialized with 
$\mathbf{u}^0=\mathbf{f}$ and $\mathbf{d}^0=D\mathbf{u}^0$.
The construction follows the idea of proximal point and 
proximal descent methods~\cite{rockafellar1976monotone,
attouch2013convergence,boct2020proximal}, but differs in 
that the proximal regularization is applied only to the 
auxiliary gradient variable $\mathbf{d}$, rather than to 
the variable pair $(\mathbf{u},\mathbf{d})$. This design preserves the original image fidelity term 
while directly controlling the variation of $\mathbf{d}$ 
between successive outer iterations.

This stabilization of $\mathbf{d}$ is essential and cannot 
be achieved by a single-loop ADMM applied directly to the 
nonconvex constrained problem~\eqref{eq:constrained}: the 
corresponding stationarity residual involves $D^\top$, 
whose nontrivial null space prevents the constraint 
residual between $D\mathbf{u}$ and $\mathbf{d}$ from 
being bounded in terms of iterate differences, which is 
a necessary condition for convergence.
Although the proximal scheme introduces a nested outer-inner 
structure, in practice a fixed modest number of inner ADMM 
iterations suffices, so the additional computational cost 
is not prohibitively expensive. The resulting proximal 
scheme is summarized in Algorithm~\ref{alg:outer-proximal}.

\begin{algorithm}[t]
\caption{TL1 gradient denoising proximal scheme}
\label{alg:outer-proximal}
\begin{algorithmic}[1]
\Require Noisy image $\mathbf{f}$; parameters $a>0$, $\mu>0$, $\eta>0$, $\beta>0$; maximum iteration $K_{\max}$; tolerance $\varepsilon_{\rm out}>0$
\State Initialize $\mathbf{u}^0\gets\mathbf{f}$, $\mathbf{d}_x^0\gets D_x\mathbf{u}^0$, $\mathbf{d}_y^0\gets D_y\mathbf{u}^0$, $\mathbf{d}^0\gets(\mathbf{d}_x^0,\mathbf{d}_y^0)$
\For{$k=0$ \textbf{to} $K_{\max}-1$}
    \State find $(\mathbf{u}^{k+1},\mathbf{d}^{k+1})$ by solving $(\mathcal{P}_\eta^k)$
    \State stop if $\|\mathbf{u}^{k+1}-\mathbf{u}^k\|_2/\max\{1,\|\mathbf{u}^k\|_2\}<\varepsilon_{\rm out}$
\EndFor
\State \Return $\mathbf{u}^{k+1}$
\end{algorithmic}
\end{algorithm}

\subsection{Properties of the TL1 Penalty} \label{subsec:tl1_properties}

We now summarize two key properties of the TL1 penalty 
$\phi_a$: weak convexity, which underpins the convergence 
analysis in Section~\ref{sec:convergence}, and a closed-form 
proximal operator, which enables an efficient $\mathbf{d}$-update 
in the ADMM inner solver.

\medskip
\paragraph{\textbf{Weak convexity.}}
We recall that a function $q:\mathbb{R}^m\to\mathbb{R}$ is said to be
$\rho$-weakly convex if
$
    \mathbf{x}\mapsto q(\mathbf{x})+\frac{\rho}{2}\|\mathbf{x}\|_2^2
$
is convex, where $\rho$ is called the weak convexity modulus. Weak convexity sits between convexity, which is too restrictive for many practical regularizers, and general nonconvexity, which offers little structure for algorithm design and convergence analysis. Intuitively, the modulus $\rho$ quantifies the amount of curvature that must be added to render a nonconvex function convex. The following
lemma gives an explicit weak convexity modulus for the TL1 penalty. The result can be traced to \cite{ahn2017difference}, where TL1 was formulated as a difference of convex function; we include a self-contained proof for completeness.

\begin{lemma}
\label{lem:tl1_weak_convexity}
For $a>0$, the scalar TL1 penalty $\phi_a(t)=\frac{(a+1)|t|}{a+|t|}$
is $\rho(a)$-weakly convex on $\mathbb{R}$ with
\begin{equation}
    \rho(a)=\frac{2(a+1)}{a^2}.
    \label{eq:rho_def}
\end{equation}
Equivalently, $t\mapsto \phi_a(t)+\frac{\rho(a)}{2}t^2$ is convex on $\mathbb{R}$.
\end{lemma}

\begin{proof}
Let
\[
    h(t):=\phi_a(t)+\frac{\rho(a)}{2}t^2.
\]
Since $\phi_a$ is even, it suffices to examine the curvature on $t>0$ and use symmetry on $t<0$. 
For $t>0$, we have
\[
    \phi_a'(t)=\frac{a(a+1)}{(a+t)^2},
    \qquad
    \phi_a''(t)
    =
    -\frac{2a(a+1)}{(a+t)^3}.
\]
Hence
\[
    \phi_a''(t)
    \ge
    -\frac{2(a+1)}{a^2}
    =
    -\rho(a),
\]
which implies
\[
    h''(t)=\phi_a''(t)+\rho(a)\ge 0,
    \qquad t>0.
\]
By symmetry, the same conclusion holds on $t<0$. 
It remains to check the behavior at the nonsmooth point $t=0$. 
The one-sided derivatives of $h$ at zero are $h'_-(0)=-\frac{a+1}{a}$, $h'_+(0)=\frac{a+1}{a}$. Thus $h'_-(0)\le h'_+(0)$. 
Therefore, the derivative of $h$ is nondecreasing across $t=0$, and since $h$ is convex on both $(-\infty,0)$ and $(0,\infty)$, we conclude that $h$ is convex on $\mathbb{R}$. 
Equivalently, $\phi_a$ is $\rho(a)$-weakly convex.
\end{proof}

\medskip
\paragraph{\textbf{Proximal operator.}}
The scalar TL1 proximal mapping is defined by
\begin{equation}
\operatorname{prox}_{\tau\phi_a}(x)
=
\arg\min_{v\in\mathbb{R}}
\left\{
    \phi_a(v)
    +
    \frac{1}{2\tau}(v-x)^2
\right\}.
\label{eq:scalar_prox_def}
\end{equation}
We present the closed-form solution for the proximal 
mapping of $\phi_a$ derived in~\cite{zhang2017minimization}.

\begin{proposition}
\label{prop:tl1_prox}
Let $a>0$ and $\tau>0$. 
For $x\in\mathbb{R}$, define
\begin{equation}
T_{\tau,a}
=
\begin{cases}
\dfrac{\tau(a+1)}{a},
&
0<\tau\le \dfrac{1}{\rho(a)},\\[1.0em]
\sqrt{2\tau(a+1)}-\dfrac{a}{2},
&
\tau> \dfrac{1}{\rho(a)},
\end{cases}
\label{eq:tl1_threshold}
\end{equation}
where $\rho(a)$ is defined in \eqref{eq:rho_def}.
Then one global minimizer of \eqref{eq:scalar_prox_def} is given by
\begin{equation}
\operatorname{prox}_{\tau\phi_a}(x)
=
\begin{cases}
0, & |x|\le T_{\tau,a},\\[0.4em]
\operatorname{sgn}(x)\,g_{\tau,a}(|x|), & |x|>T_{\tau,a},
\end{cases}
\label{eq:tl1_prox_closed}
\end{equation}
where
\begin{equation*}
g_{\tau,a}(z)
=
\frac{2}{3}(a+z)
\cos\left(
    \frac{1}{3}
    \arccos\left(
        1-\frac{27\tau a(a+1)}{2(a+z)^3}
    \right)
\right)
-\frac{2a}{3}
+\frac{z}{3},
\qquad z=|x|.
\label{eq:tl1_g}
\end{equation*}
Moreover, if $\tau<1/\rho(a)$, then the objective in
\eqref{eq:scalar_prox_def} is strongly convex; hence the global
minimizer is unique and the proximal mapping is single-valued.
\end{proposition}

For a vector $\mathbf{x}\in\mathbb{R}^m$, the TL1 penalty is defined as
\[
\Phi_a(\mathbf{x}) := \sum_{i=1}^m \phi_a(x_i).
\]
Then the proximal mapping of $\Phi_a$ is separable and is given componentwise by
\[
\operatorname{prox}_{\tau \Phi_a}(\mathbf{x})
=
\left(
\operatorname{prox}_{\tau\phi_a}(x_i)
\right)_{i=1}^m .
\]
In particular, since
$
G(\mathbf{d})=\Phi_a(\mathbf{d}_x)+\Phi_a(\mathbf{d}_y),
$
we have
\[
\operatorname{prox}_{\tau G}(\boldsymbol{\xi})
=
\left(
\operatorname{prox}_{\tau\Phi_a}(\boldsymbol{\xi}_x),
\operatorname{prox}_{\tau\Phi_a}(\boldsymbol{\xi}_y)
\right),
\qquad
\boldsymbol{\xi}=(\boldsymbol{\xi}_x,\boldsymbol{\xi}_y).
\]

\subsection{ADMM Inner Solver}\label{subsec:admm}

For each outer iteration $k$, the $k$-th subproblem in \eqref{eq:proximal_scheme} can be solved efficiently by a scaled ADMM. 
Let $\beta>0$ be the ADMM penalty parameter and let 
$\mathbf{b}=(\mathbf{b}_x,\mathbf{b}_y)$ denote the scaled dual variable for the constraint $D\mathbf{u}-\mathbf{d}=\mathbf{0}$. 
For fixed $k$, the scaled augmented Lagrangian is
\begin{equation*}
\mathcal{L}_{\beta}^{k}(\mathbf{u},\mathbf{d};\mathbf{b})
=
F(\mathbf{u})+G(\mathbf{d})
+\frac{\eta}{2}\|\mathbf{d}-\mathbf{d}^{k}\|_2^2
+\frac{\beta}{2}\|D\mathbf{u}-\mathbf{d}+\mathbf{b}\|_2^2
-\frac{\beta}{2}\|\mathbf{b}\|_2^2.
\label{eq:scaled_augmented_lagrangian}
\end{equation*}
Starting from inner variables 
$(\mathbf{u}^{k,0},\mathbf{d}^{k,0},\mathbf{b}^{k,0})$, the ADMM iterations are
\begin{equation}
\begin{cases}
\mathbf{u}^{k,\ell+1}
\in
\arg\min_{\mathbf{u}}
\mathcal{L}_{\beta}^{k}
(\mathbf{u},\mathbf{d}^{k,\ell};\mathbf{b}^{k,\ell}),
\\[0.4em]
\mathbf{d}^{k,\ell+1}
\in
\arg\min_{\mathbf{d}}
\mathcal{L}_{\beta}^{k}
(\mathbf{u}^{k,\ell+1},\mathbf{d};\mathbf{b}^{k,\ell}),
\\[0.4em]
\mathbf{b}^{k,\ell+1}
=
\mathbf{b}^{k,\ell}
+
D\mathbf{u}^{k,\ell+1}
-
\mathbf{d}^{k,\ell+1}.
\end{cases}
\label{eq:inner_admm}
\end{equation}
The dual update is the scaled dual ascent step, which enforces the consistency between the auxiliary gradient variable $\mathbf{d}$ and the discrete gradient $D\mathbf{u}$ during the ADMM iterations.

\medskip
\paragraph{\textbf{Image variable update.}}
For fixed $(\mathbf{d}^{k,\ell},\mathbf{b}^{k,\ell})$, the $\mathbf{u}$-subproblem is quadratic:
\begin{equation}
    \mathbf{u}^{k,\ell+1}
    =
    \arg\min_{\mathbf{u}}
    \frac{\mu}{2}\|\mathbf{u}-\mathbf{f}\|_2^2
    +
    \frac{\beta}{2}
    \|D\mathbf{u}-\mathbf{d}^{k,\ell}+\mathbf{b}^{k,\ell}\|_2^2.
    \label{eq:u_subproblem}
\end{equation}
The optimality condition gives
\begin{equation}
\left(\mu I+\beta D_x^\top D_x+\beta D_y^\top D_y\right)\mathbf{u}^{k,\ell+1}
=
\mu\mathbf{f}
+\beta D_x^\top\left(\mathbf{d}_x^{k,\ell}-\mathbf{b}_x^{k,\ell}\right)
+\beta D_y^\top\left(\mathbf{d}_y^{k,\ell}-\mathbf{b}_y^{k,\ell}\right).
\label{eq:u_update}
\end{equation}
Under periodic boundary conditions, $D_x$ and $D_y$ are block-circulant and can be diagonalized by the two-dimensional discrete Fourier transform. Equivalently, $D_x^\top D_x+D_y^\top D_y$ is the periodic discrete Laplacian.
Therefore, \eqref{eq:u_update} can be solved efficiently by FFT:
\begin{equation}
\mathbf{u}^{k,\ell+1}
=
\mathcal{F}^{-1}
\left(
\frac{
\mathcal{F}\left[
\mu\mathbf{f}
+
\beta D_x^\top
(\mathbf{d}_x^{k,\ell}-\mathbf{b}_x^{k,\ell})
+
\beta D_y^\top
(\mathbf{d}_y^{k,\ell}-\mathbf{b}_y^{k,\ell})
\right]
}{
\mu
+
\beta |\mathcal{F}(D_x)|^2
+
\beta |\mathcal{F}(D_y)|^2
}
\right),
\label{eq:u_update_fft}
\end{equation}
where $\mathcal{F}$ and $\mathcal{F}^{-1}$ denote the FFT 
and its inverse, respectively, and the division is performed 
componentwise in the frequency domain.
This FFT-based implementation avoids forming the full coefficient matrix and allows the linear system \eqref{eq:u_update} to be solved with complexity $O(n\log n)$, rather than the $O(n^3)$ complexity of a generic dense direct solver. 

\medskip
\paragraph{\textbf{Gradient variable update.}}

For fixed $\mathbf{u}^{k,\ell+1}$ and $\mathbf{b}^{k,\ell}$, the $\mathbf{d}$-subproblem is
\begin{equation}
\begin{aligned}
\mathbf{d}^{k,\ell+1}
=
\arg\min_{\mathbf{d}}
\quad&
G(\mathbf{d})
+
\frac{\eta}{2}\|\mathbf{d}-\mathbf{d}^k\|_2^2
+
\frac{\beta}{2}
\|D\mathbf{u}^{k,\ell+1}-\mathbf{d}+\mathbf{b}^{k,\ell}\|_2^2.
\end{aligned}
\label{eq:d_subproblem}
\end{equation}
Combining the two quadratic terms gives
\begin{equation}
\mathbf{d}^{k,\ell+1}
=
\operatorname{prox}_{\tau G}
\left(
\boldsymbol{\xi}^{k,\ell}
\right),
\qquad
\tau=\frac{1}{\beta+\eta},
\label{eq:d_update_compact}
\end{equation}
where
\begin{equation}
\boldsymbol{\xi}^{k,\ell}
=
\frac{
\beta(D\mathbf{u}^{k,\ell+1}+\mathbf{b}^{k,\ell})
+
\eta\mathbf{d}^{k}
}{
\beta+\eta
}.
\label{eq:xi_def}
\end{equation}
Equivalently,
\begin{equation}
\begin{aligned}
\boldsymbol{\xi}_x^{k,\ell}
&=
\frac{
\beta(D_x\mathbf{u}^{k,\ell+1}+\mathbf{b}_x^{k,\ell})
+
\eta\mathbf{d}_x^k
}{
\beta+\eta
},
\\
\boldsymbol{\xi}_y^{k,\ell}
&=
\frac{
\beta(D_y\mathbf{u}^{k,\ell+1}+\mathbf{b}_y^{k,\ell})
+
\eta\mathbf{d}_y^k
}{
\beta+\eta
}.
\end{aligned}
\label{eq:xi_xy_def}
\end{equation}
Since $G$ is separable, the proximal mapping in \eqref{eq:d_update_compact} is applied componentwise:
\begin{equation*}
\mathbf{d}_x^{k,\ell+1}
=
\operatorname{prox}_{\tau\Phi_a}
(\boldsymbol{\xi}_x^{k,\ell}),
\qquad
\mathbf{d}_y^{k,\ell+1}
=
\operatorname{prox}_{\tau\Phi_a}
(\boldsymbol{\xi}_y^{k,\ell}).
\label{eq:d_update_xy}
\end{equation*}

In the ADMM update, the effective proximal stepsize is 
$\tau = 1/(\beta+\eta)$ as defined in \eqref{eq:d_update_compact}. 
Thus, when $\beta+\eta > \rho(a)$, i.e., $\tau < \frac{1}{\rho(a)} 
= \frac{a^2}{2(a+1)}$, the scalar TL1 proximal objective is 
strongly convex and the gradient update is uniquely obtained 
by applying \eqref{eq:tl1_prox_closed} entrywise to 
$\boldsymbol{\xi}_x^{k,\ell}$ and $\boldsymbol{\xi}_y^{k,\ell}$.
This componentwise thresholding step plays a role analogous to soft-thresholding for the $\ell_1$ penalty. 
Unlike the convex $\ell_1$ shrinkage, however, the nonconvex TL1 penalty imposes weaker shrinkage on large gradient components, which helps preserve sharp edges and image contrast while still suppressing small oscillatory gradients caused by noise.

\medskip
\paragraph{\textbf{Dual Variable Updates.}}
The scaled dual variables are then updated to enforce consistency between the image variable and its gradient representations:
\begin{equation}
\begin{aligned}
\mathbf{b}_x^{k,\ell+1}
&=
\mathbf{b}_x^{k,\ell}
+
\big(D_x\mathbf{u}^{k,\ell+1}
-
\mathbf{d}_x^{k,\ell+1}\big),
\\
\mathbf{b}_y^{k,\ell+1}
&=
\mathbf{b}_y^{k,\ell}
+
\big(D_y\mathbf{u}^{k,\ell+1}
-
\mathbf{d}_y^{k,\ell+1}\big).
\end{aligned}
\label{eq:dual_update_final}
\end{equation}
These updates correspond to the scaled dual ascent step in ADMM and maintain the consistency between the auxiliary gradient variable $\mathbf{d}$ and the discrete gradient $D\mathbf{u}$ across the inner iterations. 
The resulting ADMM solver is summarized in Algorithm~\ref{alg:admm-solver}. 

\begin{algorithm}[t]
\caption{ADMM solver for problem \eqref{eq:proximal_scheme}}
\label{alg:admm-solver}
\begin{algorithmic}[1]
\Require Current iterate $(\mathbf{u}^k,\mathbf{d}_x^k,\mathbf{d}_y^k)$; noisy image $\mathbf{f}$; parameters $a>0$, $\mu>0$, $\eta>0$, $\beta>0$; maximum inner iterations $L_{\max}$; tolerance $\varepsilon_{\rm in}>0$
\State Initialize $\mathbf{u}^{k,0}\gets \mathbf{u}^{k}$,
       $\mathbf{d}_x^{k,0}\gets \mathbf{d}_x^{k}$,
       $\mathbf{d}_y^{k,0}\gets \mathbf{d}_y^{k}$,
       $\mathbf{b}_x^{k,0}\gets \mathbf{0}$,
       $\mathbf{b}_y^{k,0}\gets \mathbf{0}$
\State Set $\tau\gets 1/(\beta+\eta)$
\For{$\ell=0$ \textbf{to} $L_{\max}-1$}
    \State $\mathbf{u}^{\mathrm{old}}\gets \mathbf{u}^{k,\ell}$
    \State \textbf{u-update:} update $\mathbf{u}^{k,\ell+1}$ by the FFT formula \eqref{eq:u_update_fft}
    \State \textbf{d-updates:}
    \State \hspace{1em} Compute $\boldsymbol{\xi}_x^{k,\ell}$ and $\boldsymbol{\xi}_y^{k,\ell}$ by \eqref{eq:xi_xy_def}
    \State \hspace{1em} $\mathbf{d}_x^{k,\ell+1}
    \gets
    \operatorname{prox}_{\tau\Phi_a}(\boldsymbol{\xi}_x^{k,\ell})$
    \State \hspace{1em} $\mathbf{d}_y^{k,\ell+1}
    \gets
    \operatorname{prox}_{\tau\Phi_a}(\boldsymbol{\xi}_y^{k,\ell})$
    \State \textbf{dual updates:}
    \State \hspace{1em} $\mathbf{b}_x^{k,\ell+1}
    \gets
    \mathbf{b}_x^{k,\ell}
    +
    D_x\mathbf{u}^{k,\ell+1}
    -
    \mathbf{d}_x^{k,\ell+1}$
    \State \hspace{1em} $\mathbf{b}_y^{k,\ell+1}
    \gets
    \mathbf{b}_y^{k,\ell}
    +
    D_y\mathbf{u}^{k,\ell+1}
    -
    \mathbf{d}_y^{k,\ell+1}$
    \State stop if $\|\mathbf{u}^{k,\ell+1}-\mathbf{u}^{\mathrm{old}}\|_2/\max\{1,\|\mathbf{u}^{\mathrm{old}}\|_2\}<\varepsilon_{\rm in}$
\EndFor
\State \Return $(\mathbf{u}^{k,\ell+1},\mathbf{d}_x^{k,\ell+1},\mathbf{d}_y^{k,\ell+1})$
\end{algorithmic}
\end{algorithm}

\section{Convergence analysis}
\label{sec:convergence}

In this section, we analyze the convergence of the outer 
proximal scheme~\eqref{eq:proximal_scheme} summarized in 
Algorithm~\ref{alg:outer-proximal}. Specifically, we show 
that under the condition $\eta > \rho(a)$, the whole sequence 
$\{(\mathbf{u}^k, \mathbf{d}^k)\}$ generated by 
Algorithm~\ref{alg:outer-proximal} converges to a stationary 
point of the TL1 denoising model~\eqref{eq:unconstrained}. 
This is nontrivial since the overall 
problem~\eqref{eq:unconstrained} is nonconvex due to the TL1 
penalty, and standard convex ADMM convergence theory does not 
apply to the outer sequence $\{(\mathbf{u}^k, \mathbf{d}^k)\}$. 
We establish convergence via the Kurdyka--\L{}ojasiewicz (KL) 
framework~\cite{attouch2013convergence}, verifying the three 
required hypotheses of sufficient decrease, relative error, 
and continuity.

The inner ADMM iterations in Algorithm~\ref{alg:admm-solver}, 
by contrast, operate on a convex subproblem and their 
convergence is standard. For each fixed outer iterate 
$\mathbf{d}^k$, the inner subproblem $(\mathcal{P}^k_\eta)$ 
is convex when $\eta > \rho(a)$: the proximal term renders 
$G(\mathbf{d}) + \frac{\eta}{2}\|\mathbf{d}-\mathbf{d}^k\|_2^2$ 
strongly convex, and $F(\mathbf{u})$ is strongly convex by 
construction. The inner ADMM therefore falls into the classical 
two-block convex ADMM setting and converges to the unique 
solution of $(\mathcal{P}^k_\eta)$~\cite{boyd2011distributed}, 
which is treated as exact in the convergence analysis of the 
outer scheme.

We adopt standard notation from nonsmooth analysis 
\cite{rockafellar2009variational}. For a proper closed 
function $f$, we denote by $\partial f(\bar{\mathbf{x}})$ 
the \textit{limiting subdifferential} at $\bar{\mathbf{x}}$, 
which coincides with the classical convex subdifferential 
when $f$ is convex and with the gradient $\nabla f$ when 
$f$ is smooth. A point $\bar{\mathbf{x}}$ is called a 
\textit{stationary point} of $f$ if $\mathbf{0} \in 
\partial f(\bar{\mathbf{x}})$. For a nonempty closed set $C$, 
we denote by $\iota_C$ its indicator function, defined by
$
\iota_C(\mathbf{x})
=
\begin{cases}
0, & \mathbf{x}\in C,\\
+\infty, & \mathbf{x}\notin C,
\end{cases}
$
and by $N_C(\bar{\mathbf{x}}):=\partial\iota_C(\bar{\mathbf{x}})$ the limiting 
normal cone to $C$ at $\bar{\mathbf{x}}\in C$. 

\medskip
Recall that
$F(\mathbf{u})=\frac{\mu}{2}\|\mathbf{u}-\mathbf{f}\|_2^2$ and 
$G(\mathbf{d})$ defined in \eqref{eq:G_def} is the separable TL1 penalty
applied to the auxiliary gradient variable
$\mathbf{d}=(\mathbf{d}_x,\mathbf{d}_y)$. Let $\mathbf{w}:=(\mathbf{u},\mathbf{d})$,
and define the feasible set
$$
    \mathcal{C}
    :=
    \left\{
    (\mathbf{u},\mathbf{d})\in\mathbb{R}^{n}\times\mathbb{R}^{2n}
    :
    D\mathbf{u}-\mathbf{d}=\mathbf{0}
    \right\}.
$$
Now the constrained TL1 model \eqref{eq:constrained} can be written as
\begin{equation}
    \min_{\mathbf{w}}
    \;
    H(\mathbf{w})
    :=
    F(\mathbf{u})+G(\mathbf{d})+\iota_{\mathcal{C}}(\mathbf{u},\mathbf{d}).
    \label{eq:H_def}
\end{equation} 
The proximal scheme \eqref{eq:proximal_scheme} can be equivalently written as
\begin{equation}
    \mathbf{w}^{k+1}
    =
    (\mathbf{u}^{k+1},\mathbf{d}^{k+1})
    \in
    \arg\min_{\mathbf{w}}
    \left\{
        H(\mathbf{w})
        +
        \frac{\eta}{2}\|\mathbf{d}-\mathbf{d}^k\|_2^2
    \right\}.
    \label{eq:outer_iteration_compact}
\end{equation}

Next, we review the Kurdyka--\L{}ojasiewicz (KL) property, which will be used to establish convergence of the proximal sequence. For a proper lower semicontinuous function 
$f:\mathbb{R}^d\to(-\infty,+\infty]$, we denote
$
    [f<\mu]
    :=
    \{\mathbf{x}\in\mathbb{R}^d: f(\mathbf{x})<\mu\},
$
and
$
    [\gamma<f<\mu]
    :=
    \{\mathbf{x}\in\mathbb{R}^d: \gamma<f(\mathbf{x})<\mu\}.
$
Let $r_0>0$ and set
\[
\mathcal{K}(r_0):=\{\varphi:\varphi\in C^0([0,r_0))\cap C^1((0,r_0)),\ \varphi(0)=0,\ \varphi \text{ is concave, and } \varphi'>0\}.
\]
The function $f$ satisfies the KL inequality, or has the KL property, locally at
$\widetilde{\mathbf{x}}\in\operatorname{dom}\partial f$ if there exist
$r_0>0$, $\varphi\in\mathcal{K}(r_0)$, and a neighborhood
$U(\widetilde{\mathbf{x}})$ of $\widetilde{\mathbf{x}}$ such that
\begin{equation*}
    \varphi'
    \left(
        f(\mathbf{x})-f(\widetilde{\mathbf{x}})
    \right)
    \operatorname{dist}
    \left(
        \mathbf{0},
        \partial f(\mathbf{x})
    \right)
    \ge 1
    \label{eq:KL_inequality}
\end{equation*}
for all
$
    \mathbf{x}\in
    U(\widetilde{\mathbf{x}})
    \cap
    [f(\widetilde{\mathbf{x}})<f<f(\widetilde{\mathbf{x}})+r_0].
$
The function $f$ has the KL property on a set $S$ if it has the KL property at
each point of $S$.

\begin{lemma}
\label{lem:H_KL}
The function $H$ defined in \eqref{eq:H_def} is a KL function.
\end{lemma}

\begin{proof}
The function $F$ is polynomial and hence semi-algebraic. 
The scalar TL1 penalty $\phi_a(t) = \frac{(a+1)|t|}{a+|t|}$ 
is semi-algebraic, as it is a rational function of $|t|$ 
composed with the absolute value, both of which are 
semi-algebraic~\cite{attouch2010proximal}. Therefore, the separable penalty $G$ is semi-algebraic. 
Moreover, $\mathcal{C}$ is an affine subspace, so the indicator function $\iota_{\mathcal{C}}$ is also semi-algebraic. 
Hence
$
    H=F+G+\iota_{\mathcal{C}}
$
is proper, lower semicontinuous, and semi-algebraic. 
Consequently, $H$ has the KL property; see \cite{attouch2013convergence}.
\end{proof}

The convergence analysis of the sequence $\{\mathbf{w}^k\}$ 
generated by \eqref{eq:outer_iteration_compact} follows the 
descent convergence framework for KL 
functions~\cite{attouch2013convergence}, which requires 
verifying three hypotheses:

\begin{enumerate}
\item[\textbf{(H1)}] \textbf{Sufficient descent condition:}
There exists a positive constant $c_1$ such that, for all $k\in\mathbb{N}$,
\[
    c_1\|\mathbf{w}^{k+1}-\mathbf{w}^k\|_2^2
    \le
    H(\mathbf{w}^k)-H(\mathbf{w}^{k+1}).
\]

\item[\textbf{(H2)}] \textbf{Relative error condition:}
There exists a positive constant $c_2$ such that, for all $k\in\mathbb{N}$,
there exists
$\boldsymbol{\omega}^{k+1}\in\partial H(\mathbf{w}^{k+1})$ satisfying
\[
    \|\boldsymbol{\omega}^{k+1}\|_2
    \le
    c_2\|\mathbf{w}^{k+1}-\mathbf{w}^{k}\|_2.
\]

\item[\textbf{(H3)}] \textbf{Continuity condition:}
There exists a subsequence $\{\mathbf{w}^{k_j}\}_{j\in\mathbb{N}}$ and a point
$\mathbf{w}^\ast$ such that
\[
    \lim_{j\to\infty}\mathbf{w}^{k_j}=\mathbf{w}^\ast
    \quad\text{and}\quad
    \lim_{j\to\infty}H(\mathbf{w}^{k_j})=H(\mathbf{w}^\ast).
\]
\end{enumerate}

Throughout this section, we assume $\eta > \rho(a)$, so that 
$\mathbf{d} \mapsto G(\mathbf{d}) + \frac{\eta}{2}\|\mathbf{d}-\mathbf{d}^k\|_2^2$ 
is $(\eta-\rho(a))$-strongly convex by 
Lemma~\ref{lem:tl1_weak_convexity}. Together with the 
$\mu$-strong convexity of $F$, the objective 
in~\eqref{eq:outer_iteration_compact} is strongly convex on 
$\mathcal{C}$, and the iterates $\{\mathbf{w}^k\}$ are 
well-defined. For convenience, define
\begin{equation}
    \kappa := \min\{\mu,\, \eta-\rho(a)\} > 0,
    \label{eq:kappa_def}
\end{equation}
and the iterate differences
\[
    \Delta\mathbf{u}^{k+1} := \mathbf{u}^{k+1}-\mathbf{u}^k,
    \quad
    \Delta\mathbf{d}^{k+1} := \mathbf{d}^{k+1}-\mathbf{d}^k,
    \quad
    \Delta\mathbf{w}^{k+1} := (\Delta\mathbf{u}^{k+1},\Delta\mathbf{d}^{k+1}).
\]

\begin{lemma}
\label{lem:well_defined_bounded}
If $\eta > \rho(a)$, 
then the subproblem \eqref{eq:outer_iteration_compact} has a unique solution for each $k$. 
Moreover, the sequence $\{\mathbf{w}^k\}$ generated by \eqref{eq:outer_iteration_compact} is bounded.
\end{lemma}

\begin{proof}
For fixed $\mathbf{d}^k$, define
\[
    Q_k(\mathbf{u},\mathbf{d})
    :=
    H(\mathbf{u},\mathbf{d})
    +
    \frac{\eta}{2}\|\mathbf{d}-\mathbf{d}^k\|_2^2.
\]
Since $F$ is $\mu$-strongly convex and 
$\mathbf{d} \mapsto G(\mathbf{d}) + \frac{\eta}{2}
\|\mathbf{d}-\mathbf{d}^k\|_2^2$ is $(\eta-\rho(a))$-strongly 
convex by Lemma~\ref{lem:tl1_weak_convexity}, $Q_k$ is 
$\kappa$-strongly convex on $\mathcal{C}$, and hence 
\eqref{eq:outer_iteration_compact} has a unique solution 
$\mathbf{w}^{k+1}$.

Using $\mathbf{w}^k = (\mathbf{u}^k, \mathbf{d}^k)$ as a 
feasible comparison point in \eqref{eq:outer_iteration_compact}, 
we obtain
\[
    H(\mathbf{w}^{k+1})
    +
    \frac{\eta}{2}\|\mathbf{d}^{k+1}-\mathbf{d}^k\|_2^2
    \leq
    H(\mathbf{w}^k),
\]
so $\{H(\mathbf{w}^k)\}$ is nonincreasing and bounded above 
by $H(\mathbf{w}^0) < \infty$. Since every iterate satisfies $w^k=(u^k,d^k)\in C$ hence $d^k=Du^k$, and $G\ge 0$, we have
\[
    \frac{\mu}{2}\|\mathbf{u}^k-\mathbf{f}\|_2^2
    =
    F(\mathbf{u}^k)
    \leq
    F(\mathbf{u}^k) + G(\mathbf{d}^k)
    =
    H(\mathbf{w}^k)
    \leq
    H(\mathbf{w}^0),
\]
which gives boundedness of $\{\mathbf{u}^k\}$. Since 
$\mathbf{d}^k = D\mathbf{u}^k$ and $D$ is a bounded linear 
operator, $\{\mathbf{d}^k\}$ is also bounded. Hence 
$\{\mathbf{w}^k\}$ is bounded.
\end{proof}

\begin{lemma}
\label{lem:sufficient_descent}
If $\eta > \rho(a)$, then for all $k\ge0$,
\begin{equation}
    H(\mathbf{w}^k)-H(\mathbf{w}^{k+1})
    \ge
    \frac{\kappa}{2}\|\Delta\mathbf{u}^{k+1}\|_2^2
    +
    \frac{\eta+\kappa}{2}\|\Delta\mathbf{d}^{k+1}\|_2^2.
    \label{eq:sufficient_descent}
\end{equation}
Consequently, \emph{(H1)} holds with $c_1=\kappa/2$.
Moreover,
\begin{equation}
    \sum_{k=0}^{\infty}
    \|\Delta\mathbf{w}^{k+1}\|_2^2
    <\infty.
    \label{eq:square_summability}
\end{equation}
\end{lemma}

\begin{proof}
Since $Q_k$ is $\kappa$-strongly convex and $\mathbf{w}^{k+1}$ is its minimizer, we have
\[
    Q_k(\mathbf{w}^{k})
    \ge
    Q_k(\mathbf{w}^{k+1})
    +
    \frac{\kappa}{2}
    \|\mathbf{w}^{k+1}-\mathbf{w}^{k}\|_2^2.
\]
Substituting
$
    Q_k(\mathbf{w}^{k})=H(\mathbf{w}^{k})
$
and $
    Q_k(\mathbf{w}^{k+1})
    =
    H(\mathbf{w}^{k+1})
    +
    \frac{\eta}{2}\|\mathbf{d}^{k+1}-\mathbf{d}^{k}\|_2^2$,
gives \eqref{eq:sufficient_descent}. 
Since
$
    \|\Delta\mathbf{w}^{k+1}\|_2^2
    =
    \|\Delta\mathbf{u}^{k+1}\|_2^2
    +
    \|\Delta\mathbf{d}^{k+1}\|_2^2,$
\eqref{eq:sufficient_descent} implies
\[
    \frac{\kappa}{2}
    \|\Delta\mathbf{w}^{k+1}\|_2^2
    \le
    H(\mathbf{w}^{k})-H(\mathbf{w}^{k+1}).
\]
Therefore, \emph{(H1)} holds with $c_1=\kappa/2$. 
Since $H\geq 0$, summing \eqref{eq:sufficient_descent} over $k$ gives \eqref{eq:square_summability}.
\end{proof}

\begin{lemma}
\label{lem:relative_error}
If $\eta > \rho(a)$, then  for all $k\ge0$,
\begin{equation}
    \operatorname{dist}
    \left(
        0,\partial H(\mathbf{w}^{k+1})
    \right)
    \le
    \eta\|\Delta\mathbf{d}^{k+1}\|_2
    \le
    \eta\|\Delta\mathbf{w}^{k+1}\|_2.
    \label{eq:relative_error}
\end{equation}
Consequently, \emph{(H2)} holds with $c_2=\eta$.
\end{lemma}

\begin{proof}
By the optimality condition of \eqref{eq:outer_iteration_compact},
\[
    \begin{pmatrix}
        \mathbf{0}\\
        -\eta\Delta\mathbf{d}^{k+1}
    \end{pmatrix}
    \in
    \partial H(\mathbf{w}^{k+1}),
\]
which gives
\[
    \operatorname{dist}
    \bigl(
        \mathbf{0},\,\partial H(\mathbf{w}^{k+1})
    \bigr)
    \leq
    \eta\|\Delta\mathbf{d}^{k+1}\|_2
    \leq
    \eta\|\Delta\mathbf{w}^{k+1}\|_2,
\]
so \emph{(H2)} holds with $c_2 = \eta$.
\end{proof}

\begin{lemma}
\label{lem:continuity}
The sequence $\{\mathbf{w}^k\}$ admits at least one cluster point. 
Moreover, if $\mathbf{w}^{k_j}\to\mathbf{w}^\ast$, then
\begin{equation}
    H(\mathbf{w}^{k_j})\to H(\mathbf{w}^\ast).
    \label{eq:continuity_at_cluster}
\end{equation}
Consequently, \emph{(H3)} holds.
\end{lemma}

\begin{proof}
By Lemma~\ref{lem:well_defined_bounded}, the sequence $\{\mathbf{w}^k\}$ is bounded and hence has a cluster point. 
Let $\mathbf{w}^{k_j}=(\mathbf{u}^{k_j},\mathbf{d}^{k_j})\to \mathbf{w}^\ast=(\mathbf{u}^\ast,\mathbf{d}^\ast)$. 
Since each iterate is feasible, $\mathbf{d}^{k_j}=D\mathbf{u}^{k_j}$. 
Passing to the limit gives $\mathbf{d}^\ast=D\mathbf{u}^\ast$, so $\mathbf{w}^\ast\in\mathcal{C}$.

Since $F$ and $G$ are continuous on finite-dimensional spaces, we obtain
\[
    H(\mathbf{w}^{k_j})
    =
    F(\mathbf{u}^{k_j})+G(\mathbf{d}^{k_j})
    \to
    F(\mathbf{u}^{\ast})+G(\mathbf{d}^{\ast})
    =
    H(\mathbf{w}^{\ast}).
\]
Therefore \emph{(H3)} holds.
\end{proof}

We now state the main convergence theorem. 
Following the limiting-subdifferential formulation, we call $\mathbf{u}^\ast$ a stationary point of the TL1 model \eqref{eq:unconstrained} if
\begin{equation}
    0
    \in
    \nabla F(\mathbf{u}^\ast)
    +
    D^\top \partial G(D\mathbf{u}^\ast).
    \label{eq:stationarity_original_def}
\end{equation}

\begin{theorem}
\label{thm:global_convergence}
Let $\{\mathbf{w}^k\} = \{(\mathbf{u}^k, \mathbf{d}^k)\}$ 
be the sequence generated by the proximal 
scheme~\eqref{eq:outer_iteration_compact}, solved exactly 
at each iteration. If $\eta > \rho(a)$, then the following 
statements hold:
\begin{enumerate}
    \item The objective values $\{H(\mathbf{w}^k)\}$ are nonincreasing and converge to a finite limit.
    \item The sequence has finite length:
    \begin{equation}
        \sum_{k=0}^{\infty}
        \|\mathbf{w}^{k+1}-\mathbf{w}^{k}\|_2
        <\infty,
        \label{eq:finite_length}
    \end{equation}
    which implies
    $
        \|\mathbf{u}^{k+1}-\mathbf{u}^{k}\|_2\to0$ and $
        \|\mathbf{d}^{k+1}-\mathbf{d}^{k}\|_2\to0.$
        
    \item The sequence $\{(\mathbf{u}^k,\mathbf{d}^k)\}$ converges to a point $(\mathbf{u}^\ast,\mathbf{d}^\ast)$.
    \item The limit point $(\mathbf{u}^\ast,\mathbf{d}^\ast)$ is a stationary point of the constrained problem \eqref{eq:constrained}. 
    Consequently, $\mathbf{u}^\ast$ is a stationary point of the TL1 denoising model \eqref{eq:unconstrained} in the sense of \eqref{eq:stationarity_original_def}.
\end{enumerate}
\end{theorem}

\begin{proof}
By Lemma~\ref{lem:sufficient_descent}, the sequence $\{H(\mathbf{w}^k)\}$ is nonincreasing. 
Since $H$ is bounded below by $0$, $\{H(\mathbf{w}^k)\}$ converges to a finite limit. 
Moreover, Lemma~\ref{lem:sufficient_descent} verifies \emph{(H1)}, Lemma~\ref{lem:relative_error} verifies \emph{(H2)}, and Lemma~\ref{lem:continuity} verifies \emph{(H3)}.

Since $H$ is a KL function by Lemma~\ref{lem:H_KL} and the sequence generated by \eqref{eq:outer_iteration_compact} satisfies hypotheses \emph{(H1)}--\emph{(H3)}, the KL convergence theorem \cite{attouch2013convergence} for descent sequences implies the finite length property
$
    \sum_{k=0}^{\infty}
    \|\mathbf{w}^{k+1}-\mathbf{w}^{k}\|_2
    <\infty.
$
By Lemma~\ref{lem:well_defined_bounded}, $\{\mathbf{w}^k\}$ is bounded. 
Together with the finite length property, this implies that the whole sequence converges to a single limit point, denoted by
$
    \mathbf{w}^\ast=(\mathbf{u}^\ast,\mathbf{d}^\ast).
$

It remains to identify the stationarity condition. 
By Lemma~\ref{lem:relative_error} and the finite length property, there exists
$\boldsymbol{\omega}^{k+1}\in \partial H(\mathbf w^{k+1})$ such that
$
    \|\boldsymbol{\omega}^{k+1}\|_2\to 0.
$
Moreover, since $\mathbf w^k\to \mathbf w^\ast$ and Lemma~\ref{lem:continuity}
gives the continuity of $H$ along convergent subsequences of $\{\mathbf w^k\}$,
we have
$
    H(\mathbf w^k)\to H(\mathbf w^\ast).
$
Consequently, $\mathbf w^{k+1}\to \mathbf w^\ast$, $H(\mathbf w^{k+1})\to H(\mathbf w^\ast)$, and $\boldsymbol{\omega}^{k+1}\to \mathbf 0$. Since $\boldsymbol{\omega}^{k+1}\in \partial H(\mathbf w^{k+1})$ and the limiting subdifferential is sequentially closed under convergence of both the points and the corresponding function values, we obtain
$
    0\in\partial H(\mathbf w^\ast).
$

We next verify the stationarity condition. The subdifferential 
sum rule gives
\[
    \partial H(\mathbf{w}^\ast)
    \subset
    \begin{pmatrix}
        \nabla F(\mathbf{u}^\ast)\\
        \partial G(\mathbf{d}^\ast)
    \end{pmatrix}
    +
    N_{\mathcal{C}}(\mathbf{u}^\ast,\mathbf{d}^\ast),
\]
where $N_{\mathcal{C}}(\mathbf{u}^\ast,\mathbf{d}^\ast)$ 
denotes the limiting normal cone to $\mathcal{C}$ at 
$(\mathbf{u}^\ast,\mathbf{d}^\ast)$. Since $\mathcal{C}$ 
is a linear subspace, this coincides with its orthogonal 
complement:
\[
    N_{\mathcal{C}}(\mathbf{u}^\ast,\mathbf{d}^\ast)
    =
    \mathcal{C}^\perp
    =
    \left\{
        (D^\top\mathbf{y},\,-\mathbf{y})
        :
        \mathbf{y}\in\mathbb{R}^{2n}
    \right\}.
\]
From $\mathbf{0}\in\partial H(\mathbf{w}^\ast)$, there exist
$\mathbf{y}^\ast\in\mathbb{R}^{2n}$ and
$\mathbf{z}^\ast\in\partial G(\mathbf{d}^\ast)$ such that
\[
    \mathbf{0}
    =
    \begin{pmatrix}
        \nabla F(\mathbf{u}^\ast)\\
        \mathbf{z}^\ast
    \end{pmatrix}
    +
    \begin{pmatrix}
        D^\top\mathbf{y}^\ast\\
        -\mathbf{y}^\ast
    \end{pmatrix}.
\]

Since $\mathbf{w}^\ast \in \mathcal{C}$, we have 
$D\mathbf{u}^\ast = \mathbf{d}^\ast$, so the two equations 
above give
\begin{equation}
\begin{cases}
    \mathbf{0} = \nabla F(\mathbf{u}^\ast) 
    + D^\top\mathbf{y}^\ast,\\[0.3em]
    \mathbf{y}^\ast \in \partial G(\mathbf{d}^\ast),\\[0.3em]
    D\mathbf{u}^\ast = \mathbf{d}^\ast,
\end{cases}
\label{eq:stationarity_constrained}
\end{equation}
which are precisely the stationarity conditions for the 
constrained problem \eqref{eq:constrained}. Substituting 
$\mathbf{d}^\ast = D\mathbf{u}^\ast$ into the first two 
lines yields
$
    \mathbf{0}
    \in
    \nabla F(\mathbf{u}^\ast)
    +
    D^\top\partial G(D\mathbf{u}^\ast),
$
which is \eqref{eq:stationarity_original_def}. 
This completes the proof.
\end{proof}

\begin{remark}[Single-loop alternative]
\label{rem:single_loop}
An alternative single-loop ADMM with adaptive penalty 
$\beta_k = \beta_0\sigma^k$, $1 < \sigma < 3$, can be 
analyzed following the framework of~\cite{mcpYou2018}, 
adapted to the TL1 penalty. However, this yields only 
subsequential convergence to a stationary point, which 
is weaker than Theorem~\ref{thm:global_convergence}. Furthermore, 
the adaptive penalty $\beta_k \to \infty$ may cause 
numerical instability for large $k$. We therefore adopt 
the proximal scheme as the primary algorithm, which 
provides whole sequence convergence under the mild 
condition $\eta > \rho(a)$ with a fixed penalty 
parameter $\beta$.
\end{remark}

\section{Numerical experiments}

\begin{figure}[!t]
    \centering
    \begin{subfigure}{0.26\columnwidth}
        \includegraphics[width=\linewidth]{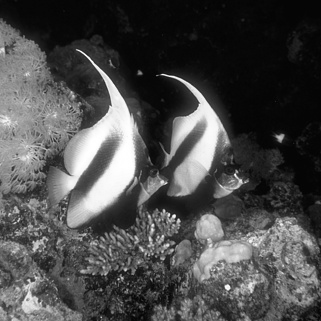}
        \caption{Fish}
    \end{subfigure}\hfill
    \begin{subfigure}{0.26\columnwidth}
        \includegraphics[width=\linewidth]{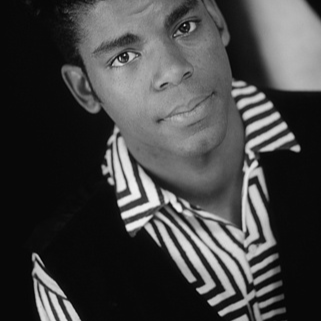}
        \caption{Striped Shirt}
    \end{subfigure}\hfill
    \begin{subfigure}{0.26\columnwidth}
        \includegraphics[width=\linewidth]{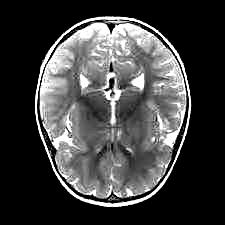}
        \caption{MRI}
    \end{subfigure}\\[3pt]
    \begin{subfigure}{0.26\columnwidth}
        \includegraphics[width=\linewidth]{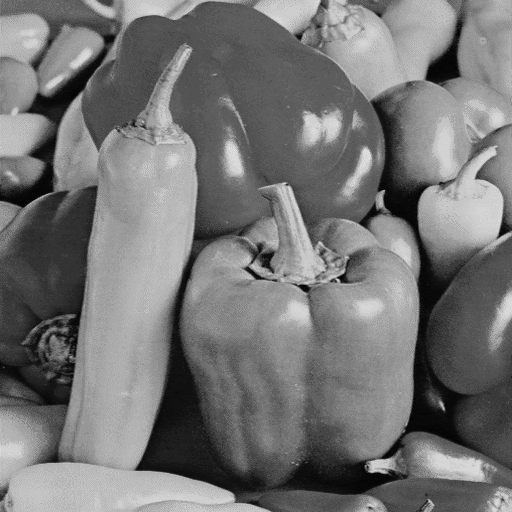}
        \caption{Peppers}
    \end{subfigure}\hfill
    \begin{subfigure}{0.26\columnwidth}
        \includegraphics[width=\linewidth]{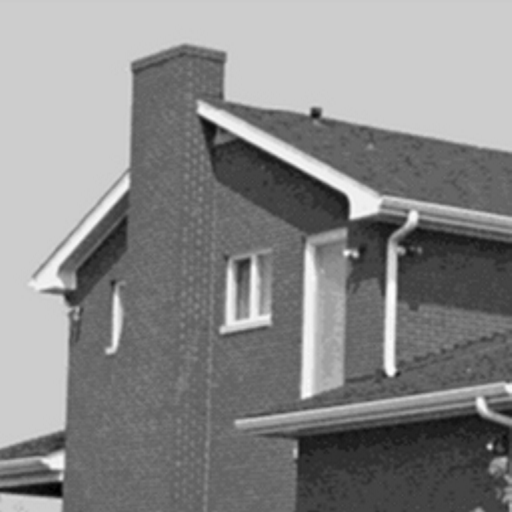}
        \caption{House}
    \end{subfigure}\hfill
    \begin{subfigure}{0.26\columnwidth}
        \includegraphics[width=\linewidth]{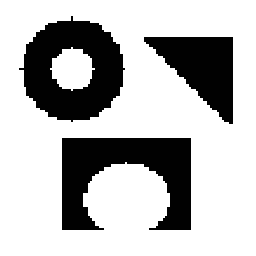}
        \caption{Shapes}
    \end{subfigure}
    \caption{Clean Benchmark Images.}
    \label{fig:original_images}
\end{figure}

\begin{table}[!t]
\centering
\scriptsize
\begin{tabular}{l c l}
\toprule
\textbf{Image} & \textbf{Size} & \textbf{Role in evaluation} \\
\midrule
\textit{Shapes} & $256\times256$ & piecewise-constant regions and sharp edges \\
\textit{Fish} & $321\times321$ & natural texture and local contrast \\
\textit{Striped Shirt} & $321\times321$ & high-frequency stripe patterns and facial shading \\
\textit{Peppers} & $512\times512$ & smooth surfaces, highlights, and object boundaries \\
\textit{House} & $512\times512$ & repeated architectural edges and brick texture \\
\textit{MRI} & $255\times255$ & structured grayscale image \\
\bottomrule
\end{tabular}
\caption{Test images used in the experiments.}
\label{tab:test_images}
\end{table}

This section presents numerical experiments demonstrating the effectiveness of the proposed model and its practical implementation. We compare the TL1-based method against several gradient-based regularization models: TV~\cite{rudin1992nonlinear}, $\ell_1$--$\ell_2$~\cite{YifeiAni-iso2015}, MCP~\cite{mcpYou2018}, and a combination of a logarithmic function with high-order TV~\cite{FAN202585}, referred to as LOG+TV. 

Six benchmark images were used in the experiments. 
\textit{Shapes} serves as a simple image to evaluate each model's baseline performance. 
\textit{Fish} and \textit{Striped Shirt} correspond to images 62 and 65, respectively, from the BSD68 subset of the Berkeley Segmentation Dataset~\cite{martin2001database,zhang2017beyond}, providing a mix of patterns and textures for the models to preserve; both are cropped to $321 \times 321$ pixels. 
\textit{Peppers} and \textit{House} are standard test images from the Set12 denoising benchmark~\cite{zhang2017beyond}.
We also include a brain MRI image from the Brain Tumor MRI Dataset on Kaggle~\cite{msoud_nickparvar_2026}, referred to as \textit{MRI}.
The clean images are shown in Figure~\ref{fig:original_images}, and their dimensions are provided in Table~\ref{tab:test_images}. All images were normalized to $[0,1]$ before adding zero-mean Gaussian noise with standard deviation $\sigma\in\{0.01,0.03,0.10\}$. 

To assess the denoising performance, we use Peak Signal-to-Noise Ratio (PSNR) \cite{hore2010psnr} and Structural Similarity Index Measure (SSIM) \cite{wang2004ssim}. PSNR is defined as: 
\begin{equation*}\label{PSNR}
\text{PSNR}(X,Y) = 10\log_{10}\left(
\frac{R^2}{\frac{1}{NM}\sum_{i=1}^N\sum_{j=1}^M (X_{i,j}-Y_{i,j})^2}
\right),
\end{equation*}
where $R$ is the maximum pixel value, $X$ is the original image, and $Y$ is the denoised image. SSIM is defined locally as:  
\begin{equation*}
\text{ssim}(x,y) = \frac{(2\mu_x\mu_y + c_1)(2\sigma_{xy} + c_2)}{(\mu_x^2 + \mu_y^2 + c_1)(\sigma_x^2 + \sigma_y^2 + c_2)},
\end{equation*}
Then the overall SSIM is computed as: 
\begin{equation*}
\text{SSIM}(X,Y) = \frac{1}{P}\sum_{i=1}^P \text{ssim}(x_i,y_i),
\end{equation*}
where $\mu_x, \mu_y$ are local means, $\sigma_x^2, \sigma_y^2$ are local variances, $\sigma_{xy}$ is covariance, and $c_1, c_2$ are stability constants (set to $c_1=(0.01)^2$, $c_2=(0.03)^2$ in all experiments). Higher PSNR and SSIM values indicate better denoising performance. PSNR reflects pixel-wise fidelity, and SSIM measures the preservation of structural information such as edges and local contrast.

All algorithms were implemented in MATLAB R2024a and run on a Windows laptop with a 13th Gen Intel Core i7-1355U processor and 16 GB RAM. For each method, the tunable parameters were selected by Bayesian optimization~\cite{snoek2012practical} using MATLAB's \texttt{bayesopt} function, with SSIM as the optimization objective. The iterations were terminated when the relative change between successive reconstructions fell below $10^{-4}$ or when the maximum number of outer iterations $K_{\max}=200$ was reached. For TL1, each proximal subproblem was solved by at most $L_{\max}=10$ ADMM iterations. 

\subsection{Quantitative results}

\begin{table}[!t]
\centering
\scriptsize
\setlength{\tabcolsep}{2.5pt}
\renewcommand{\arraystretch}{1.18}
\caption{Quantitative comparison of denoising performance. Each entry reports SSIM / PSNR, with the best result in each row highlighted in bold.}
\label{tab:all_metrics}

\begin{adjustbox}{max width=\textwidth}
\begin{tabular}{@{}llccccc@{}}
\toprule
\textbf{Image}
& $\boldsymbol{\sigma}$
& \textbf{TV}
& \textbf{$\ell_1$--$\ell_2$}
& \textbf{MCP}
& \textbf{LOG+TV}
& \textbf{TL1} \\
\midrule

\multirow{3}{*}{\textit{Fish}}
& 0.01
& 0.9864 / \textbf{39.49}
& 0.9844 / 37.03
& 0.9660 / 34.93
& 0.8965 / 27.14
& \textbf{0.9873} / 38.75 \\
& 0.03
& 0.9429 / 31.83
& 0.9430 / 31.83
& 0.9521 / 32.40
& 0.8957 / 27.33
& \textbf{0.9565} / \textbf{32.97} \\
& 0.10
& 0.8412 / 26.76
& 0.8407 / 26.74
& \textbf{0.8417} / 26.72
& 0.8018 / 25.32
& 0.8415 / \textbf{26.77} \\

\midrule
\multirow{3}{*}{\textit{Striped Shirt}}
& 0.01
& 0.9892 / 36.81
& 0.9868 / 39.85
& 0.9863 / 39.21
& 0.9726 / 30.05
& \textbf{0.9920} / \textbf{41.23} \\
& 0.03
& 0.9765 / 33.57
& \textbf{0.9811} / 35.17
& 0.9808 / \textbf{36.53}
& 0.9630 / 29.58
& 0.9746 / 33.25 \\
& 0.10
& 0.8925 / 28.38
& 0.8924 / 28.38
& 0.9059 / 28.10
& \textbf{0.9176} / 27.75
& 0.9126 / \textbf{28.46} \\

\midrule
\multirow{3}{*}{\textit{MRI}}
& 0.01
& 0.9889 / 35.96
& 0.9876 / 34.35
& 0.9861 / 38.00
& 0.9056 / 22.52
& \textbf{0.9914} / \textbf{39.93} \\
& 0.03
& 0.9421 / 29.80
& 0.9547 / 32.23
& \textbf{0.9559} / 31.91
& 0.8830 / 22.63
& 0.9554 / \textbf{32.75} \\
& 0.10
& 0.7604 / 25.17
& 0.7605 / \textbf{25.18}
& 0.7461 / 23.61
& 0.7110 / 21.57
& \textbf{0.7608} / 25.15 \\

\midrule
\multirow{3}{*}{\textit{Peppers}}
& 0.01
& \textbf{0.9823} / 40.48
& 0.9822 / \textbf{40.56}
& 0.9279 / 34.63
& 0.9266 / 32.37
& \textbf{0.9823} / 40.55 \\
& 0.03
& 0.9295 / 33.69
& 0.9276 / 33.78
& 0.9212 / 33.65
& 0.9129 / 31.73
& \textbf{0.9302} / \textbf{34.10} \\
& 0.10
& 0.8285 / 28.99
& 0.8277 / 28.96
& 0.8395 / 27.61
& 0.8584 / 28.83
& \textbf{0.8694} / \textbf{29.67} \\

\midrule
\multirow{3}{*}{\textit{House}}
& 0.01
& 0.9918 / 42.47
& 0.9865 / 40.77
& 0.9911 / \textbf{43.46}
& 0.9904 / 38.16
& \textbf{0.9920} / 42.96 \\
& 0.03
& \textbf{0.9753} / 37.93
& \textbf{0.9753} / 37.88
& 0.9576 / 36.22
& 0.9751 / 36.05
& \textbf{0.9753} / \textbf{37.94} \\
& 0.10
& 0.8600 / 30.55
& 0.8611 / 30.56
& 0.9358 / \textbf{32.78}
& \textbf{0.9378} / 32.34
& 0.9297 / 31.90 \\

\midrule
\multirow{3}{*}{\textit{Shapes}}
& 0.01
& 0.9970 / 42.02
& 0.9993 / 47.49
& 0.9988 / 46.76
& 0.9597 / 25.47
& \textbf{0.9995} / \textbf{48.70} \\
& 0.03
& 0.9870 / 36.41
& 0.9886 / 36.93
& 0.9902 / 37.92
& 0.9541 / 25.39
& \textbf{0.9919} / \textbf{38.68} \\
& 0.10
& 0.9096 / 27.30
& 0.9128 / 27.35
& 0.9242 / 25.62
& 0.8865 / 22.78
& \textbf{0.9281} / \textbf{28.10} \\

\bottomrule
\end{tabular}
\end{adjustbox}
\end{table}

Table~\ref{tab:all_metrics} reports SSIM and PSNR for all methods, images, and noise levels. The results show that TL1 is competitive across all benchmark images and is particularly strong on images with sparse, high-contrast gradient structures. 
On \textit{Shapes}, TL1 achieves the best PSNR and SSIM 
at all three noise levels, indicating that the TL1 penalty 
effectively removes noise from flat regions while preserving 
sharp boundaries. TL1 also performs strongly on 
\textit{Peppers} and \textit{Fish}, attaining the best 
PSNR and SSIM at $\sigma=0.03$ and $\sigma=0.10$ on the 
former, and at $\sigma=0.03$ on the latter.

The $\ell_1$--$\ell_2$ and MCP models perform well on 
textured or mid-noise cases, while LOG+TV attains high 
SSIM under strong noise. These variations are expected, as different nonconvex penalties encode distinct trade-offs among noise suppression, edge preservation, and texture smoothing.  Overall, TL1 is particularly effective for images with sparse, high-contrast gradient structures while remaining competitive on more textured images. 

\subsection{Qualitative results}

For visual comparison, each example is displayed in three rows: the restored images with a selected region marked by a green box, the corresponding enlarged region, and the difference map. 

\begin{figure}[!t]
    \centering
    \scriptsize
    \setlength{\tabcolsep}{1pt}
    \begin{tabular}{c c c c c c}
        \textbf{Noisy} & \textbf{TV} & \textbf{$\ell_1$--$\ell_2$} & \textbf{MCP} & \textbf{LOG+TV} & \textbf{TL1} \\[1mm]

        \includegraphics[width=0.155\textwidth]{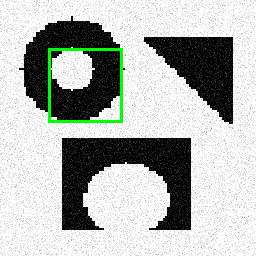} &
        \includegraphics[width=0.155\textwidth]{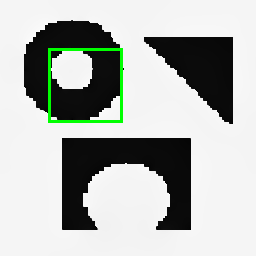} &
        \includegraphics[width=0.155\textwidth]{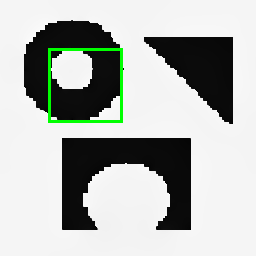} &
        \includegraphics[width=0.155\textwidth]{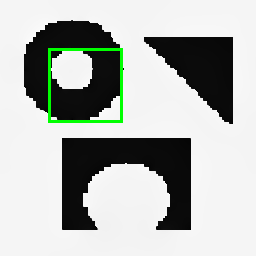} &
        \includegraphics[width=0.155\textwidth]{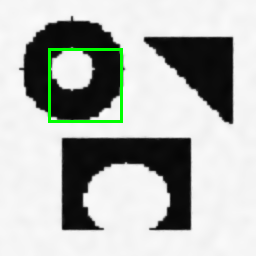} &
        \includegraphics[width=0.155\textwidth]{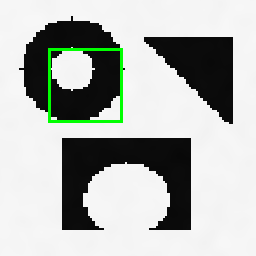} \\[1mm]

        \includegraphics[width=0.155\textwidth]{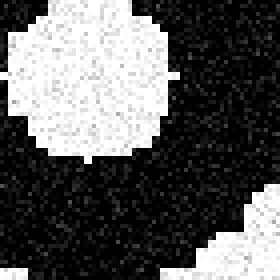} &
        \includegraphics[width=0.155\textwidth]{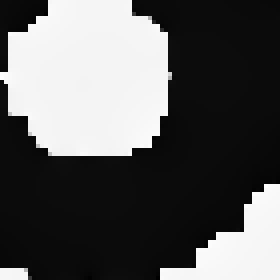} &
        \includegraphics[width=0.155\textwidth]{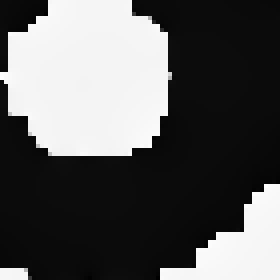} &
        \includegraphics[width=0.155\textwidth]{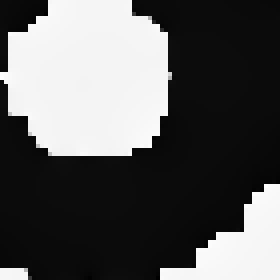} &
        \includegraphics[width=0.155\textwidth]{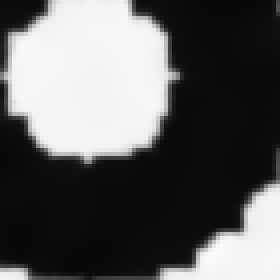} &
        \includegraphics[width=0.155\textwidth]{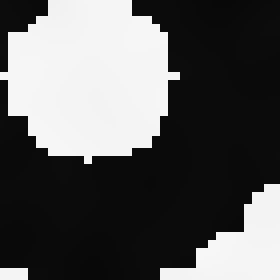} \\[1mm]

        \includegraphics[width=0.155\textwidth]{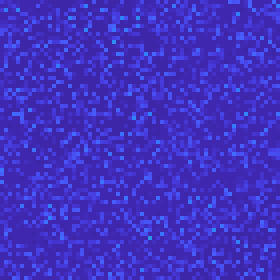} &
        \includegraphics[width=0.155\textwidth]{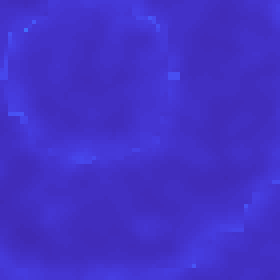} &
        \includegraphics[width=0.155\textwidth]{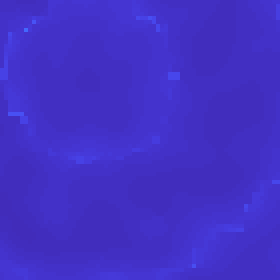} &
        \includegraphics[width=0.155\textwidth]{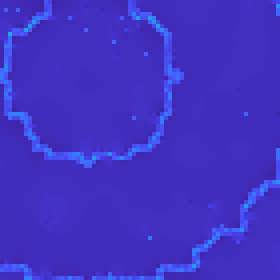} &
        \includegraphics[width=0.155\textwidth]{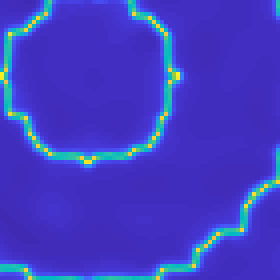} &
        \includegraphics[width=0.155\textwidth]{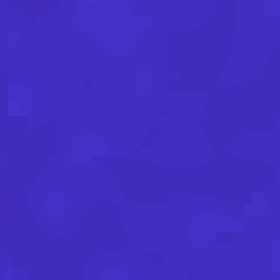}
    \end{tabular}
    \caption{Qualitative comparison for \textit{Shapes} at $\sigma=0.10$. Rows from top to bottom: full images with selected regions, zoomed-in regions, and zoomed-in difference maps.}
    \label{fig:shape_zoom_results}
\end{figure}

Figure~\ref{fig:shape_zoom_results} shows the \textit{Shapes} image at noise level $\sigma=0.10$. This image is piecewise constant, so the main visual differences arise near sharp transitions between black and white regions. In the enlarged region, TV and $\ell_1$--$\ell_2$ remove most of the noise but slightly round the boundary. MCP preserves a sharper transition, yet its difference map reveals visible errors around the edge. LOG+TV produces an overly smoothed boundary with contrast loss near the selected region. By comparison, TL1 recovers clean flat regions while maintaining sharp edges, consistent with its best SSIM and PSNR values for \textit{Shapes} in Table~\ref{tab:all_metrics}.

\begin{figure}[!t]
    \centering
    \captionsetup[subfigure]{labelformat=parens,labelsep=space}
    \begin{subfigure}[c]{0.22\textwidth}
        \centering
        \includegraphics[width=\linewidth]{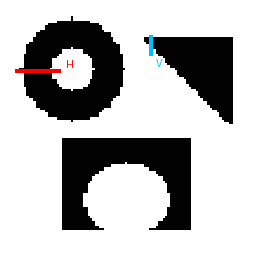}
        \caption{Profile locations}
    \end{subfigure}
    \hfill
    \begin{subfigure}[c]{0.74\textwidth}
        \centering
        \includegraphics[width=\linewidth]{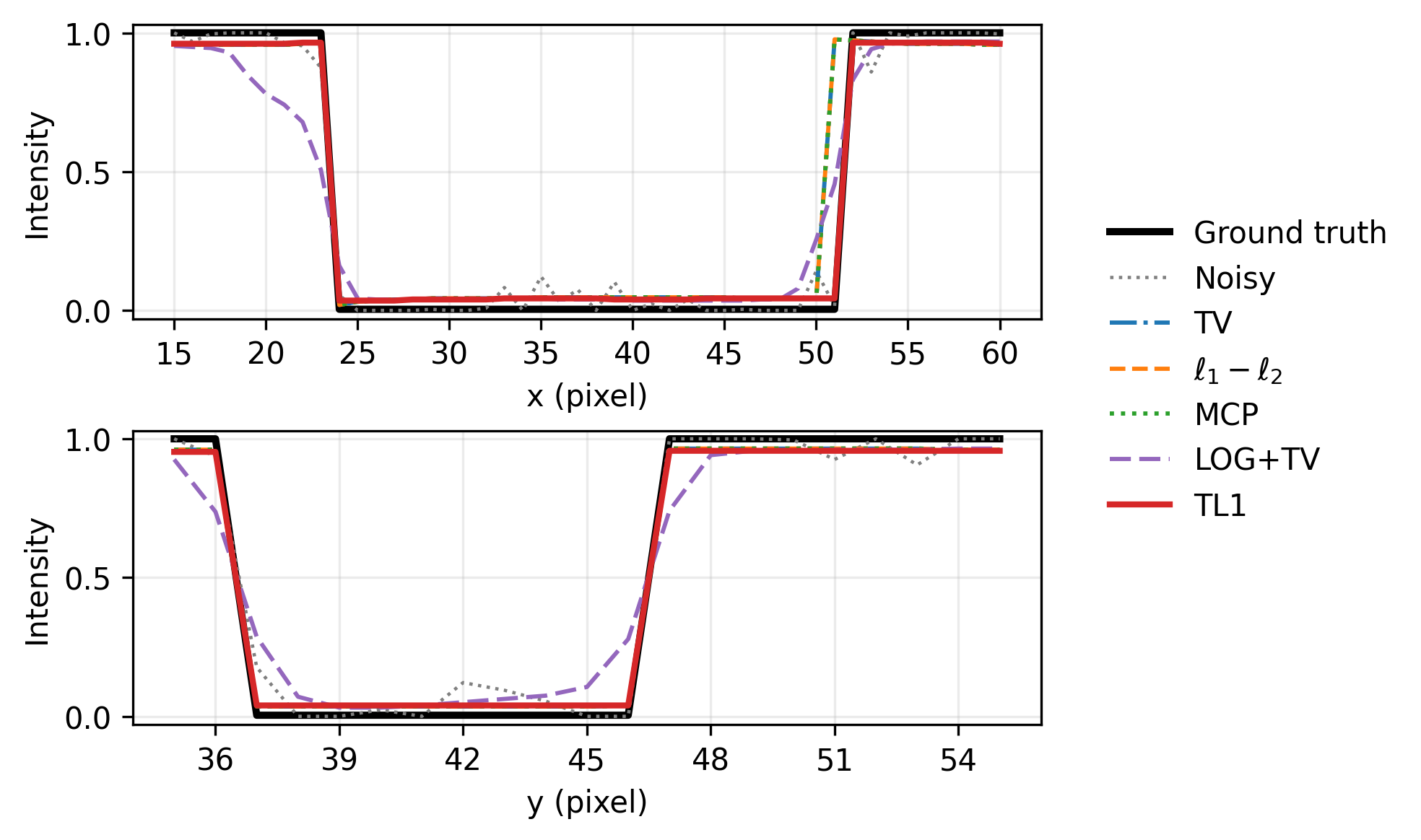}
        \caption{Local profiles}
    \end{subfigure}
    \caption{Local line profiles for \textit{Shapes} at $\sigma=0.10$. The red horizontal segment is taken along row $y=70$ with $x\in[15,60]$, and the cyan vertical segment along column $x=150$ with $y\in[35,55]$.}
    \label{fig:shape_line_profiles}
\end{figure}

To further quantify edge preservation, Figure~\ref{fig:shape_line_profiles} reports local line profiles across two sharp intensity transitions in \textit{Shapes}. TL1 closely follows the ground-truth step profiles, preserving both the flat plateaus and abrupt edge locations, and LOG+TV spreads the transitions over a wider region and introduces visible contrast loss near the edges.

\begin{figure}[!t]
    \centering
    \scriptsize
    \setlength{\tabcolsep}{1pt}
    \begin{tabular}{c c c c c c}
        \textbf{Noisy} & \textbf{TV} & \textbf{$\ell_1$--$\ell_2$} & \textbf{MCP} & \textbf{LOG+TV} & \textbf{TL1} \\[1mm]

        \includegraphics[width=0.155\textwidth]{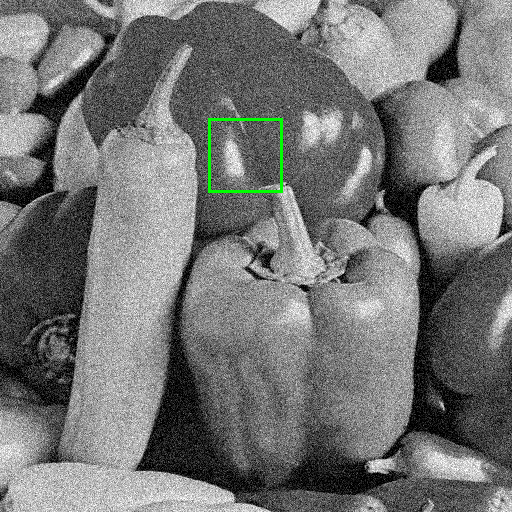} &
        \includegraphics[width=0.155\textwidth]{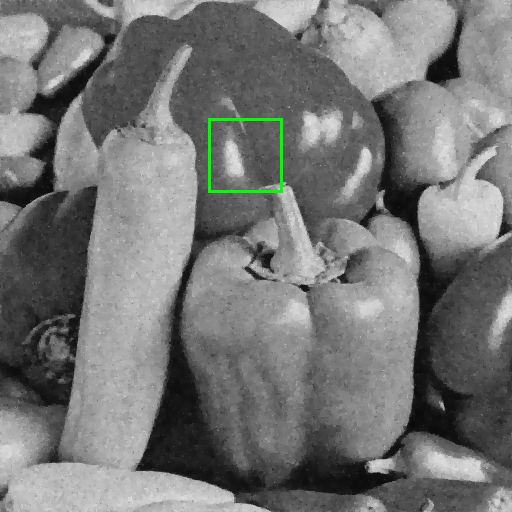} &
        \includegraphics[width=0.155\textwidth]{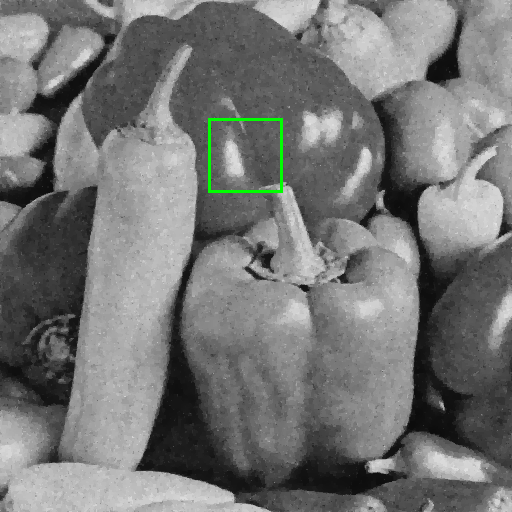} &
        \includegraphics[width=0.155\textwidth]{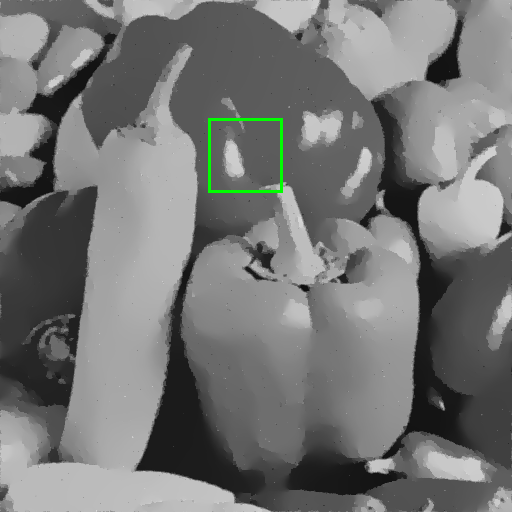} &
        \includegraphics[width=0.155\textwidth]{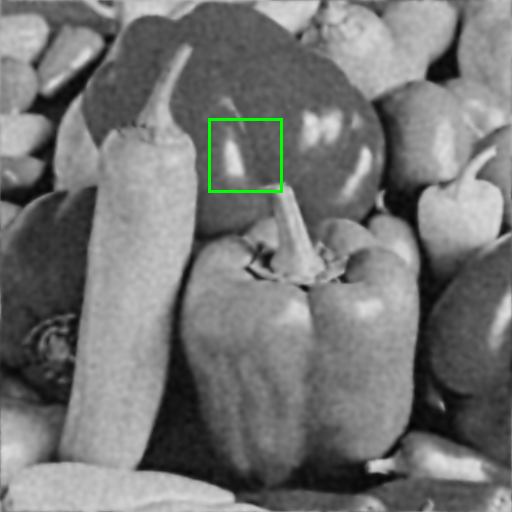} &
        \includegraphics[width=0.155\textwidth]{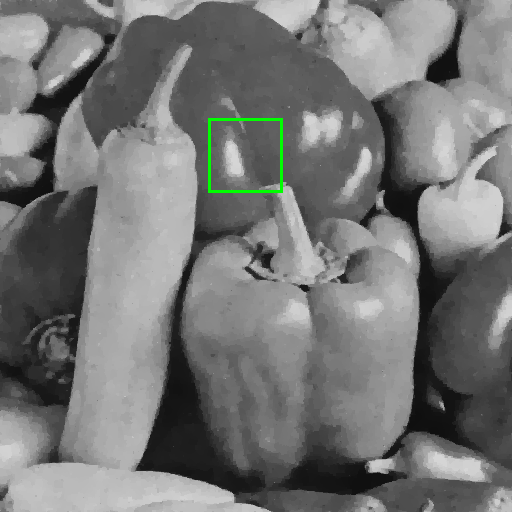} \\[1mm]

        \includegraphics[width=0.155\textwidth]{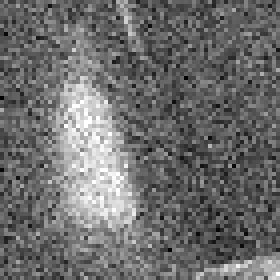} &
        \includegraphics[width=0.155\textwidth]{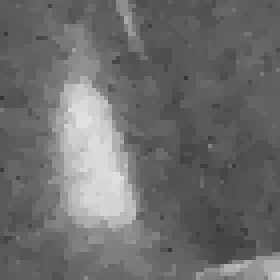} &
        \includegraphics[width=0.155\textwidth]{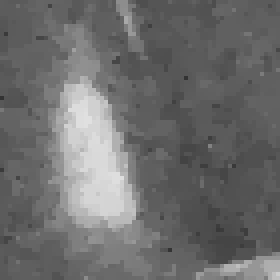} &
        \includegraphics[width=0.155\textwidth]{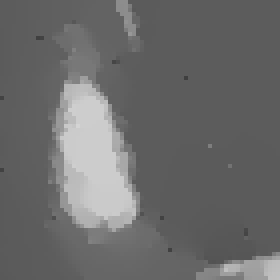} &
        \includegraphics[width=0.155\textwidth]{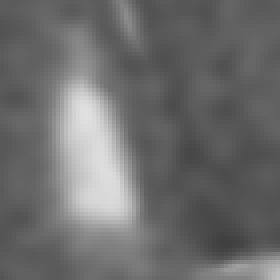} &
        \includegraphics[width=0.155\textwidth]{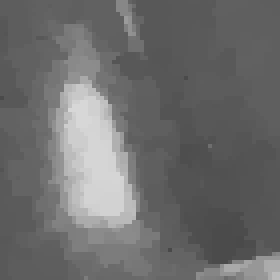} \\[1mm]

        \includegraphics[width=0.155\textwidth]{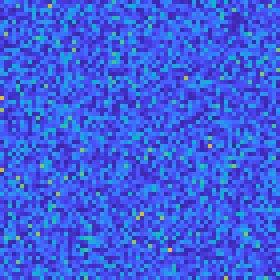} &
        \includegraphics[width=0.155\textwidth]{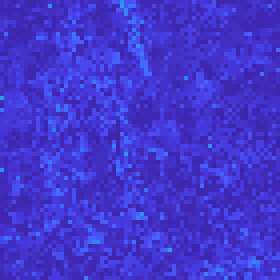} &
        \includegraphics[width=0.155\textwidth]{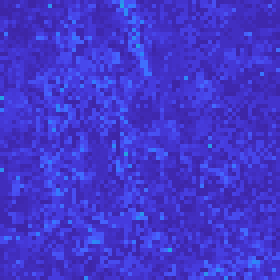} &
        \includegraphics[width=0.155\textwidth]{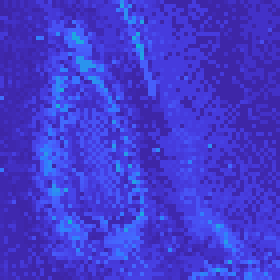} &
        \includegraphics[width=0.155\textwidth]{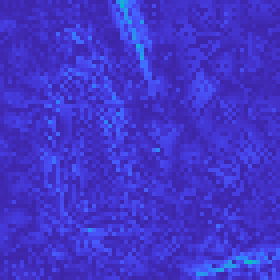} &
        \includegraphics[width=0.155\textwidth]{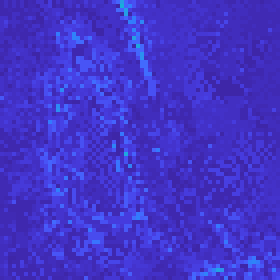}
    \end{tabular}
    \caption{Qualitative comparison for \textit{Peppers} at $\sigma=0.10$. Rows from top to bottom: full images with selected regions, zoomed-in regions, and zoomed-in difference maps.}
    \label{fig:peppers_zoom_results}
\end{figure}

Figure~\ref{fig:peppers_zoom_results} shows the \textit{Peppers} image at noise level $\sigma=0.10$, a more challenging case than \textit{Shapes} due to the coexistence of smooth regions, fine details, and sharp edges. 
The zoomed region contains smooth intensity gradients, specular highlights, and object boundaries under strong noise. TV and $\ell_1$--$\ell_2$ exhibit visible residual noise, while MCP and LOG+TV over-smooth local contrast. TL1 produces a cleaner reconstruction with the main boundary and intensity transition better preserved. The difference maps confirm that the reconstruction error of TL1 is more spatially concentrated and less diffuse in the selected region. 

\begin{figure}[!t]
    \centering
    \scriptsize
    \setlength{\tabcolsep}{1pt}
    \begin{tabular}{c c c c c c}
        \textbf{Noisy} & \textbf{TV} & \textbf{$\ell_1$--$\ell_2$} & \textbf{MCP} & \textbf{LOG+TV} & \textbf{TL1} \\[1mm]

        \includegraphics[width=0.155\textwidth]{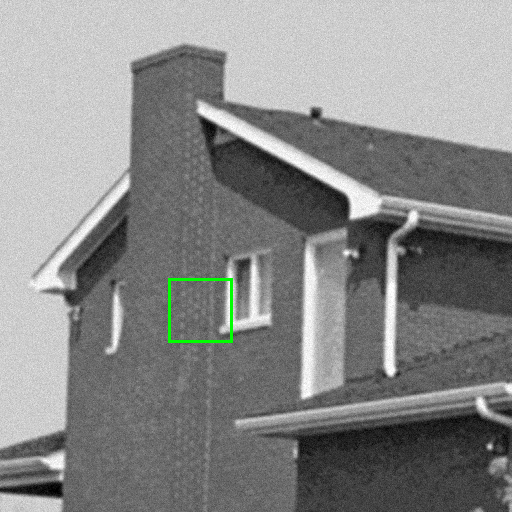} &
        \includegraphics[width=0.155\textwidth]{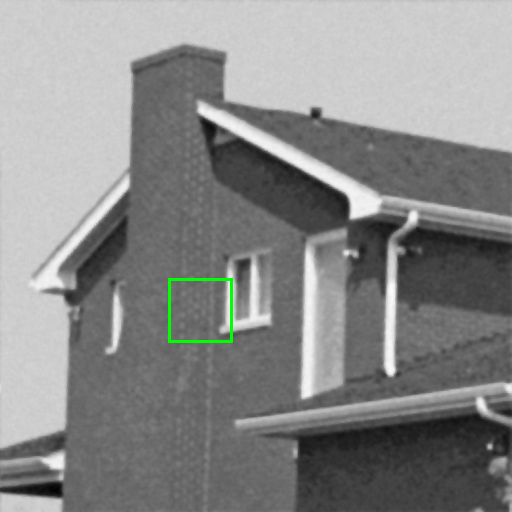} &
        \includegraphics[width=0.155\textwidth]{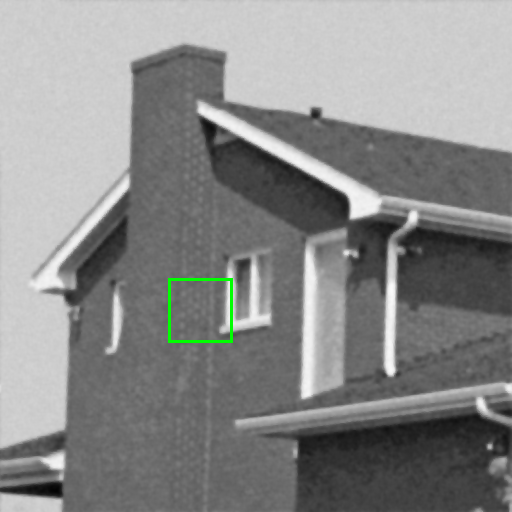} &
        \includegraphics[width=0.155\textwidth]{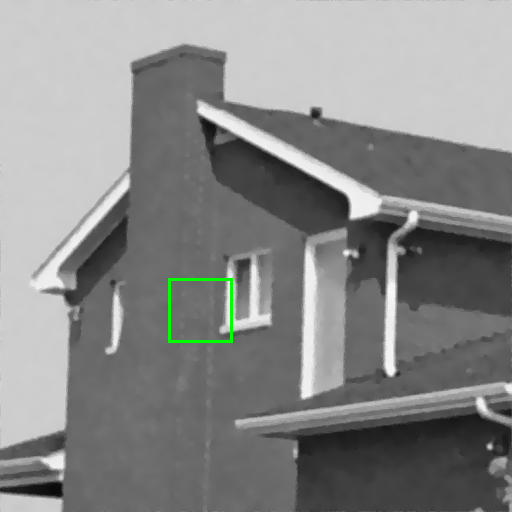} &
        \includegraphics[width=0.155\textwidth]{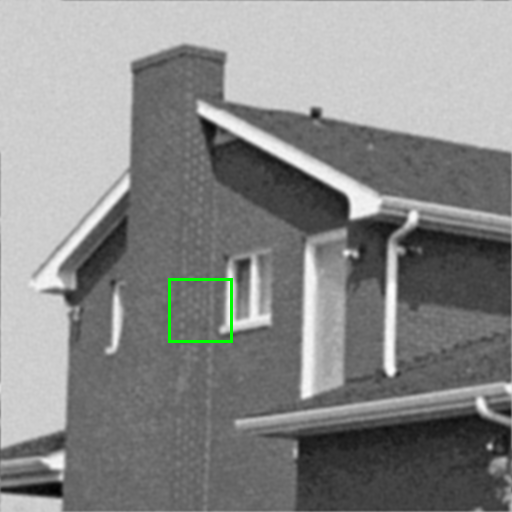} &
        \includegraphics[width=0.155\textwidth]{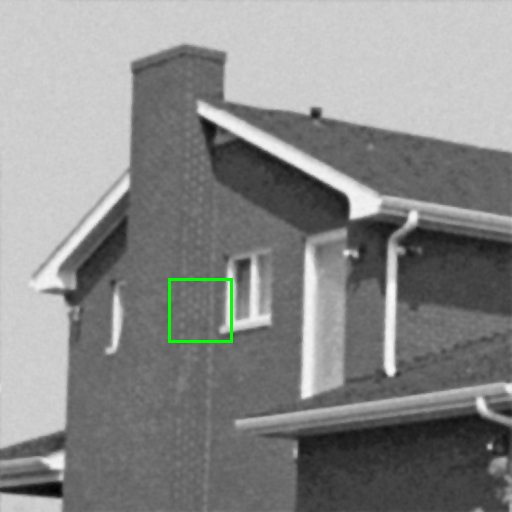} \\[1mm]

        \includegraphics[width=0.155\textwidth]{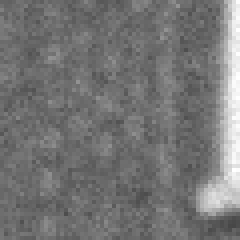} &
        \includegraphics[width=0.155\textwidth]{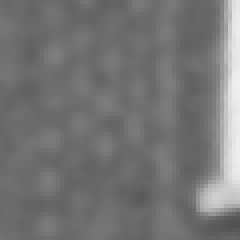} &
        \includegraphics[width=0.155\textwidth]{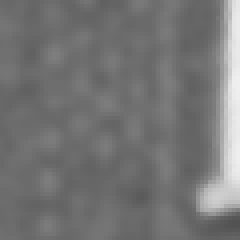} &
        \includegraphics[width=0.155\textwidth]{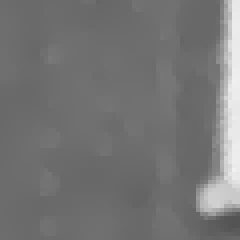} &
        \includegraphics[width=0.155\textwidth]{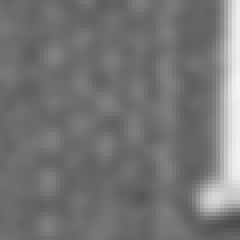} &
        \includegraphics[width=0.155\textwidth]{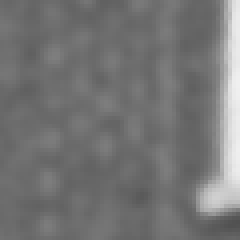} \\[1mm]

        \includegraphics[width=0.155\textwidth]{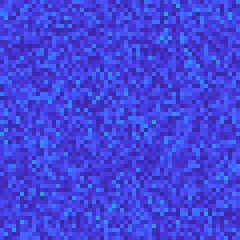} &
        \includegraphics[width=0.155\textwidth]{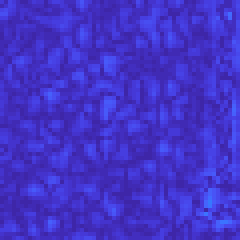} &
        \includegraphics[width=0.155\textwidth]{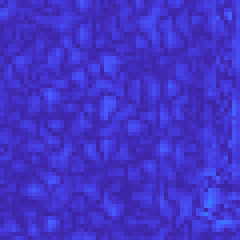} &
        \includegraphics[width=0.155\textwidth]{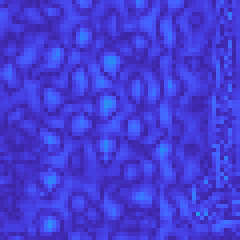} &
        \includegraphics[width=0.155\textwidth]{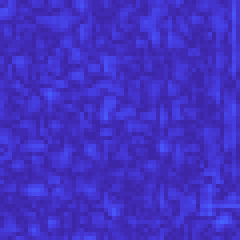} &
        \includegraphics[width=0.155\textwidth]{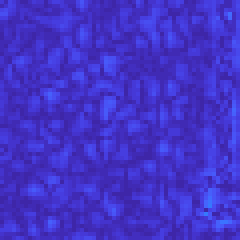}
    \end{tabular}
    \caption{Qualitative comparison for \textit{House} at $\sigma=0.03$. Rows from top to bottom: full images with selected regions, zoomed-in regions, and zoomed-in difference maps.}
    \label{fig:house_zoom_results}
\end{figure}

Figure~\ref{fig:house_zoom_results} shows the \textit{House} image at noise level $\sigma=0.03$. The selected region for zoom-in contains repeated brick structures and sharp local edges. In the enlarged region, MCP loses part of the brick pattern, and LOG+TV produces weaker local textures. TV, $\ell_1$--$\ell_2$, and TL1 achieve comparable (global) PSNR and SSIM values, yet TL1 better preserves the local brick structure, as confirmed by the difference maps.

\begin{figure}[!t]
    \centering
    \scriptsize
    \setlength{\tabcolsep}{1pt}
    \begin{tabular}{c c c c c c}
        \textbf{Noisy} & \textbf{TV} & \textbf{$\ell_1$--$\ell_2$} & \textbf{MCP} & \textbf{LOG+TV} & \textbf{TL1} \\[1mm]

        \includegraphics[width=0.155\textwidth]{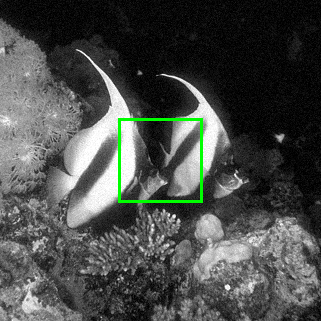} &
        \includegraphics[width=0.155\textwidth]{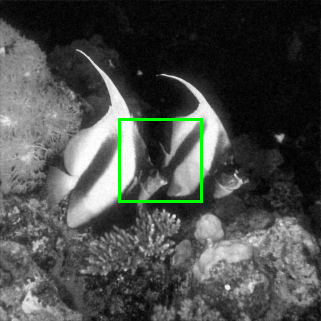} &
        \includegraphics[width=0.155\textwidth]{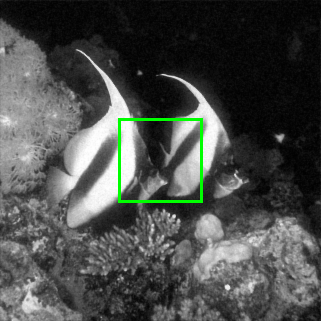} &
        \includegraphics[width=0.155\textwidth]{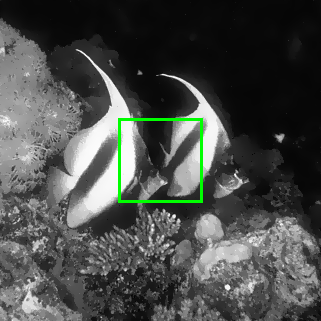} &
        \includegraphics[width=0.155\textwidth]{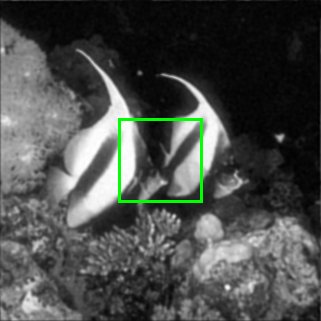} &
        \includegraphics[width=0.155\textwidth]{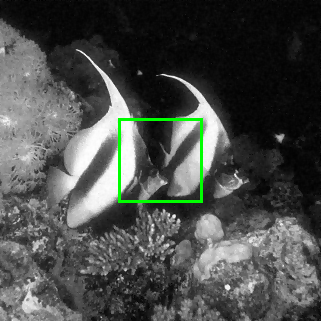} \\[1mm]

        \includegraphics[width=0.155\textwidth]{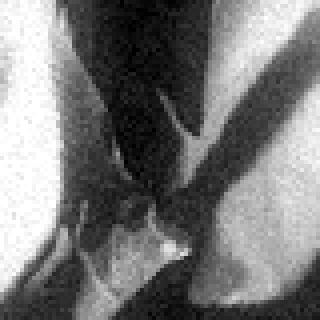} &
        \includegraphics[width=0.155\textwidth]{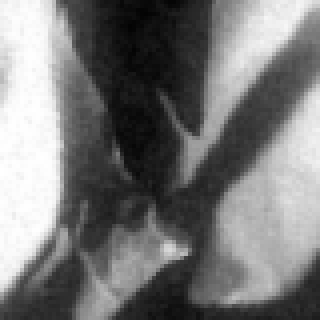} &
        \includegraphics[width=0.155\textwidth]{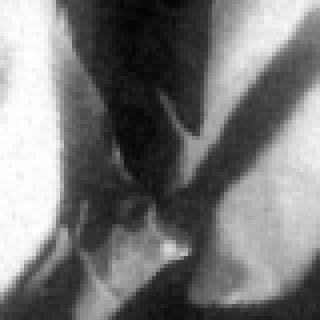} &
        \includegraphics[width=0.155\textwidth]{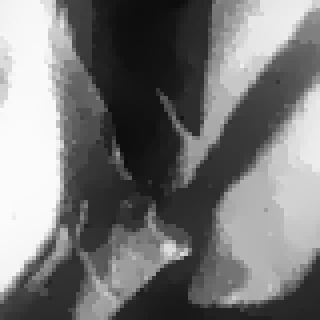} &
        \includegraphics[width=0.155\textwidth]{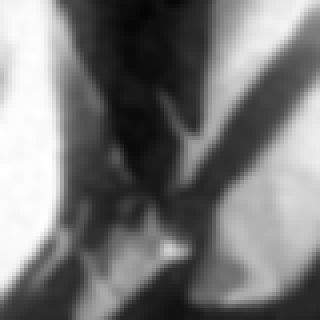} &
        \includegraphics[width=0.155\textwidth]{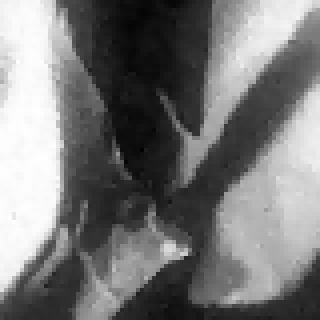} \\[1mm]

        \includegraphics[width=0.155\textwidth]{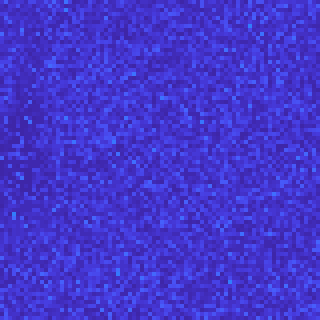} &
        \includegraphics[width=0.155\textwidth]{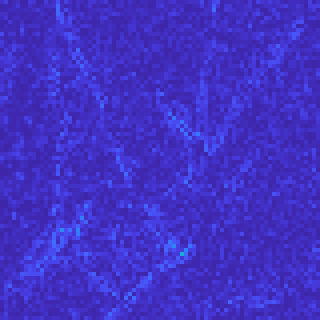} &
        \includegraphics[width=0.155\textwidth]{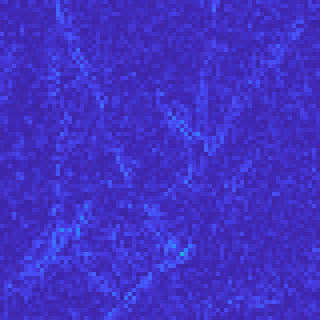} &
        \includegraphics[width=0.155\textwidth]{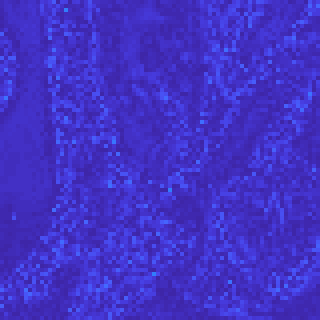} &
        \includegraphics[width=0.155\textwidth]{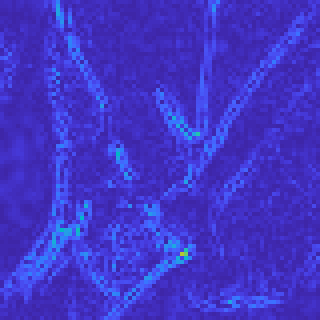} &
        \includegraphics[width=0.155\textwidth]{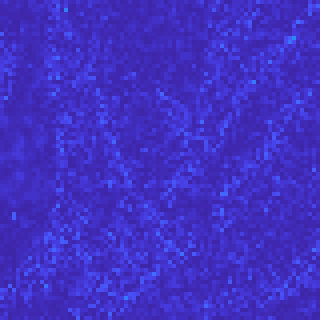}
    \end{tabular}
    \caption{Qualitative comparison for \textit{Fish} at $\sigma=0.03$. Rows from top to bottom: full images with selected regions, zoomed-in regions, and zoomed-in difference maps.}
    \label{fig:fish_zoom_results}
\end{figure}

Figure~\ref{fig:fish_zoom_results} shows the \textit{Fish} image at noise level $\sigma=0.03$. The selected region contains smooth shading, local contrast variations, and fine structural transitions. In the enlarged region, MCP and LOG+TV either blur local structures or retain visible residual error. TL1 better preserves transitions between neighboring structures with fewer artifacts. The difference map is correspondingly more spatially concentrated.

\section{Conclusion}

We proposed a TL1 gradient regularization model for image denoising and solved the resulting nonsmooth nonconvex problem using a proximal outer scheme with an ADMM inner solver. The TL1 penalty provides a flexible alternative to TV by promoting sparse gradients while reducing the shrinkage bias on large edge components. The algorithm benefits from an FFT-based image update and a closed-form componentwise TL1 proximal step, and the proximal iterates are shown to converge to a stationary point under a suitable weak-convexity condition. Numerical experiments demonstrate that TL1 is competitive with different regularization methods across multiple Gaussian noise levels. The method is effective on piecewise-smooth and high-contrast images, where it preserves sharp boundaries and local contrast while suppressing noise in homogeneous regions. Future work will extend this framework to other imaging inverse problems, including deblurring, inpainting, and super-resolution, and will investigate adaptive parameter selection strategies.

\section*{Acknowledgments}
N.C. was partially supported by the University of North Carolina at Chapel Hill Summer Undergraduate Research Fellowships (SURF) Program. J.J. was partially supported by AMS-Simons Travel Grant 330934.

\bibliographystyle{unsrt}
\bibliography{ref}

\end{document}